\documentclass[10pt,5p,twocolumn]{elsarticle}

\usepackage[english]{babel}

\usepackage{graphicx}
\usepackage[font=footnotesize]{caption}
\usepackage{subcaption}
\usepackage{booktabs}
\usepackage{amsmath,amssymb}
\usepackage{slashed}

\usepackage{axodraw4j}
\usepackage{sectsty}
\allsectionsfont{\sffamily\mdseries\upshape}

\allowdisplaybreaks

\usepackage[unicode=true,ps2pdf,colorlinks]{hyperref}
\hypersetup{pdftitle="Two-Hadron Saturation for the Pseudoscalar-Vector-Vector Correlator and Phenomenological Applications"}

\begin{document}

\begin{frontmatter}

\title{Two-Hadron Saturation for the Pseudoscalar-Vector-Vector Correlator and Phenomenological Applications}
\author[prg]{Tom\'{a}\v{s} Husek}\ead{husek@ipnp.mff.cuni.cz}
\author[upp]{Stefan Leupold}\ead{stefan.leupold@physics.uu.se}

\address[prg]{Institute of Particle and Nuclear Physics, Faculty of Mathematics and Physics, Charles University\\
V Hole\v{s}ovi\v{c}k\'{a}ch 2, Praha 8, Czech Republic\vspace{2mm}}
\address[upp]{Institutionen f\"or Fysik och Astronomi, Uppsala Universitet, Box 516, S--75120 Uppsala, Sweden}

\begin{abstract}
The pseudoscalar-vector-vector ($PVV$) correlator is constructed using two meson multiplets in the vector and two in the pseudoscalar channel.
The parameters are constrained by the operator product expansion at leading order where two or all three momenta are considered as large.
Demanding in addition the Brodsky--Lepage limit one obtains (in the chiral limit) a pion-vector-vector ($\pi VV$) correlator with only one free parameter.
The singly virtual pion transition form factor $\mathcal F_{\pi^0\gamma\gamma^\ast}$ and the decay width of $\omega\to\pi^0\gamma$ are independent of this parameter and can serve as cross-checks of the results.
The free parameter is determined from a fit of the $\omega$-$\pi$ transition form factor $\mathcal F_{\pi^0\omega\gamma^{*}}$.
The resulting $\pi VV$ correlator is used to calculate the decay widths $\omega\to\pi^0e^+e^-$ and $\omega\to\pi^0\mu^+\mu^-$ and finally the widths of the rare decay $\pi^0\to e^+e^-$ and of the Dalitz decay $\pi^0\to e^+e^-\gamma$.
Incorporating radiative QED corrections the calculations of $\pi^0$ decays are compared to the KTeV results.
We find a deviation of 2 $\sigma$ or less for the rare pion decay.
\end{abstract}

\begin{keyword}
13.20.Cz Decays of $\pi$ mesons \sep 13.40.Gp Electromagnetic form factors \sep 11.30.Rd Chiral symmetry \sep 12.40.Vv Vector-meson dominance
\end{keyword}

\end{frontmatter}


\section{Introduction and summary}
\label{sec:intro}

Two major challenges of contemporary particle physics are the search for beyond-Standard Model physics and a better understanding of the non-perturbative low-energy sector of the strong interaction.
Typically both aspects intermix when it comes to high-precision determinations of low-energy quantities and the corresponding Standard Model predictions.
If a low-energy observable is potentially influenced by quantum effects from new particles, then it is also influenced by hadronic loop effects.
The latter often --- if not always --- constitute the main uncertainty of the Standard Model prediction.
On the other hand, new physics can only be revealed if a significant deviation between experiment and Standard Model calculation is observed.
Of course, this requires small uncertainties for both the experimental and the Standard Model result.
Our poor understanding of non-perturbative QCD could provide a serious hurdle for a reliable Standard Model calculation and/or for a reliable uncertainty estimate of such a calculation.

Two quantities of current interest which might indicate some deviation between experiment and the Standard Model are the gyromagnetic ratio of the muon~\cite{Jegerlehner:2009ry,Czerwinski:2012ry} and the rare decay of the neutral pion into electron and positron~\cite{Abouzaid:2006kk,Dorokhov:2007bd}.
An important quantity that enters both observables is the pion transition form factor, i.e.\ the three-point correlator between a neutral pion and two electromagnetic currents.

Two tasks are carried out in the present work: First, the pion-vector-vector ($\pi VV$) correlator is determined by combining high-energy quark-based information with low(er)-energy hadronic information.
We follow the general approach proposed in~\cite{Knecht:2001xc} with some refinements.
Second, we explore some phenomenological consequences of our correlator and focus in particular on the rare pion decay $\pi^0 \to e^+ e^-$.
Here we include also QED radiative corrections along the lines of~\cite{Vasko:2011pi,Husek:2014tna}.

Two limits of QCD are of particular interest for low-energy hadron physics: The chiral limit where the masses of the two or three lightest quarks are neglected~\cite{Weinberg:1978kz,Wess:1971yu,Witten:1983tw,Gasser:1983yg,Gasser:1984gg} and the limit where the number of quark colors, $N_c$, is sent to infinity~\cite{'tHooft:1973jz,Witten:1979kh}.
Concerning the $\pi VV$ correlator the chiral anomaly fixes the low-energy strength unambiguously in the chiral limit.
For a large number of colors there are infinitely many, infinitely narrow, i.e.\ stable, quark-antiquark states for every combination of quantum numbers.
They show up as poles in the $n$-point correlators of quark currents.
In the real world of three colors the hadrons generically turn to unstable resonances because the hadronic interactions do not vanish any more.
The poles in the correlators turn to cuts (and poles in other Riemann sheets).
The cuts start at the corresponding many-body thresholds.
Thus the relevance of the large-$N_c$ limit for the real world is highest, if one considers hadrons which are narrow and/or kinematical regions where there are no (significant) cuts.
This is our guiding principle when exploring phenomenological consequences.
Concerning form factors there are no cuts in the space-like region.

In the Standard Model the rare decay of the pion into electron and positron is caused by a loop where the pion first turns into a pair of (real or virtual) photons; see Fig.~\ref{fig:piee} below.
At this vertex the pion transition form factor sneaks in.
If the form factor was replaced by a constant, the loop would diverge~\cite{Drell:1959nc,Savage:1992ac,Knecht:1999gb}.
In QCD the pion transition form factor is suppressed for large virtualities~\cite{Lepage:1979zb,Lepage:1980fj,Brodsky:1981rp}.
This leads to a finite result for the $\pi^0$-$e^+$-$e^-$ amplitude at the one-loop level of the Standard Model calculation.
Thus for a quantitative determination of the branching ratio of the considered rare pion decay it is necessary to know where and how fast the pion transition form factor reaches its asymptotic form which in turn depends on the various combinations of virtualities.

These considerations show that one needs information from various QCD regimes: The threshold regime governed by the chiral anomaly, the regime of hadronic resonances, and, finally, the regime of asymptotically high energies dictated by quarks and 
asymptotic freedom.
These regimes are connected in the approach of~\cite{Knecht:2001xc} where the operator product expansion (OPE) for various three-point correlators of quark currents is matched to a hadronic ansatz that satisfies the chiral constraints for the low-energy limit.
This ansatz is furthermore based on a truncation of the infinite tower of stable hadronic states that appears in the limit of a large number of colors.

In principle one can work out arbitrary many orders in the OPE and match to the parameters that emerge with the tower of hadron states.
However, the higher orders in the OPE contain unknown quark and gluon condensates of high dimensionality.
Thus in practice the model dependence emerges from a selection of the to be matched OPE constraints and from the choice where to truncate the tower of hadron states.
Using one hadron multiplet per channel and leading-order OPE constraints has been studied in detail in~\cite{Knecht:2001xc}.
This truncation is called ``lowest-meson dominance'' (LMD).
In the present work we will explore the consequences of having two hadron multiplets per channel.
In the language of~\cite{Knecht:2001xc} our approach would be called ``LMD+V+P''.
To avoid this clumsy name we decided to introduce the name ``two-hadron saturation'' (THS).

Concerning our quantity of interest, the $\pi VV$ correlator and the corresponding pion transition form factor, the starting point on the level of quark currents is the pseudoscalar-vector-vector ($PVV$) correlator.
The consequences of LMD for this quantity have been studied in~\cite{Knecht:2001xc}.
An application to the rare pion decay to electron and positron was presented in~\cite{Knecht:1999gb}.
Two vector multiplets have also been considered in~\cite{Knecht:2001xc,Mateu:2007tr}; see also~\cite{Dorokhov:2007bd} where this has been used for the rare pion decay.
What makes our approach different from previous works is that we explore in detail the consequence of two multiplets in {\em each} channel and that we fit and/or compare to data on the $\pi^0\omega V$ correlator.
In fact, including a second multiplet in the vector channel involves the energy region of about 1.4 GeV \cite{PDG}.
In this region there is also a pseudoscalar multiplet.
Thus the extension from LMD to two multiplets for a channel suggests to use two multiplets for {\em every} channel.
Concerning the second aspect, the interrelation to the $\pi^0\omega V$ correlator, we will come back to this issue below, after discussing in more detail the pertinent high-energy constraints.

There is yet one more short-distance limit to be considered.
Instead of studying the high-energy limit of correlators of quark currents (OPE), one can also study the high-energy limit of correlators that involve specific asymptotic states like hadrons or photons together with one or several quark currents.
In particular the high-energy behavior of the pion-photon-vector correlator has recently gained much attention since the BaBar data~\cite{Aubert:2009mc} seem to contradict the Brodsky--Lepage (B-L) scaling limit~\cite{Brodsky:1981rp} while the Belle data~\cite{Uehara:2012ag} seem to support it.

Using THS we are able to satisfy all leading-order OPE constraints for the $PVV$ correlator and in addition the B-L constraint for the pion-photon-vector correlator.
While LMD satisfies the same OPE constraints, it violates the B-L constraint as can be easily deduced from the explicit form given in~\cite{Knecht:1999gb}; see also the discussion in~\cite{Knecht:2001xc}.
As we will show below, the constraints from the leading order of the OPE, from B-L and from the chiral anomaly together fix the THS approach to the $\pi VV$ correlator up to one single parameter, which we call $\kappa$.
If the invariant mass of one of the vector currents in this correlator is set to zero, i.e.\ for the pion-photon-vector correlator, then $\kappa$ drops out.
In other words we have full predictive power for this correlator.
Aiming at the rare pion decay into electron and positron one needs the full information on the $\pi VV$ correlator for arbitrary invariant masses of the two vector currents.
In principle, the parameter $\kappa$ would be best determined from data on the $\pi VV$ correlator with both invariant masses being different from real photons.
Unfortunately such data do not exist.

In this situation we turn to the second-best choice.
Projecting the $\pi VV$ correlator on one of the vector mesons that we include in THS yields a 3-point correlator for the pion, the vector meson and a quark current with vector quantum numbers.
Given that our approach is based on the large-$N_c$ limit where mesons are approximated by infinitely narrow states, it is suggestive to use a narrow vector meson.
Since phenomenologically and in the large-$N_c$ limit the pion decouples from the $\phi$ meson~\cite{Okubo:1963fa,Iizuka:1966fk,Zweig:1981pd,Witten:1979kh}, we are left with the $\omega$ meson as the best choice for a vector meson.
Consequently we will use data on the $\omega$-$\pi$ transition form factor~\cite{Arnaldi:2009aa} to fix our remaining parameter $\kappa$.

The $\pi VV$ correlator obtained in this way from THS shows an awesome behavior: If the virtuality of one vector current becomes large while the other is set to zero (photon case), the asymptotic B-L limit is reached rather fast, resembling essentially strict vector-meson dominance (VMD)~\cite{sakuraiVMD,Landsberg:1986fd}.
The scale that defines where the approach to the asymptotic limit sets in is basically given by the mass of the $\rho$/$\omega$ meson.
On the other hand, if both vector currents have the same large virtuality, the corresponding asymptotic limit is reached very late for the $\pi VV$ correlator as obtained from THS.
This finding points to the relevance of details of hadronic physics above 1 GeV.
For this case the $\pi VV$ correlator from VMD falls off much faster than demanded by QCD, while for LMD the asymptotic limit is reached much earlier than for THS.
Since the rare decay $\pi^0 \to e^+ e^-$ is sensitive to both high- and low-energy physics, it is interesting to study how this intriguing behavior of the $\pi VV$ correlator obtained from THS influences the branching ratio of this rare process.

Before going into the details of our findings we shall compare our approach to related ones from the literature.
The $\omega$-$\pi$ transition form factor and related quantities have also been addressed in~\cite{Terschluesen:2010ik,Terschlusen:2012xw,Terschlusen:2013iqa} where the ground-state vector mesons are treated as light degrees of freedom.
By construction the approach is restricted to low energies, i.e.\ high-energy constraints were not considered.
Nonetheless, it turns out that the THS result for the $\omega$-$\pi$ transition form factor is numerically very close to the one from~\cite{Terschluesen:2010ik}.

Conceptually close in spirit to THS is the Lagrangian approach utilized, e.g., in~\cite{RuizFemenia:2003hm,Roig:2013baa} (earlier references can be traced back from these works).
Also here hadron resonances in the large-$N_c$ limit are used to interpolate between the low-energy region governed by chiral perturbation theory and the high-energy region governed by the OPE or quark scaling considerations.
In~\cite{Roig:2013baa} a $PVV$ and a corresponding $\pi VV$ correlator are constructed with one multiplet in the vector channel and two multiplets in the pseudoscalar channel (i.e.\ the Goldstone bosons and one resonance multiplet).
For an extension to two multiplets in the vector channel see~\cite{Roig:2014uja}.
The $\pi VV$ correlator from~\cite{Roig:2013baa} satisfies the B-L constraint, but the $PVV$ correlator satisfies only one leading-order OPE constraint, not all of them.
As shown in~\cite{Kampf:2011ty} the Lagrangian utilized in \cite{Roig:2013baa} is not capable to provide correlators that satisfy all OPE constraints; see also the discussion in~\cite{Knecht:2001xc}.
Instead of the other OPE constraints on the $PVV$ correlator a high-energy constraint based on quark counting rules is imposed on the $\pi \rho V$ correlator in~\cite{Roig:2013baa}.

In our work we impose high-energy constraints on the $PVV$ correlator (THS satisfies {\em all} leading-order OPE constraints) and on the $\pi VV$ correlator (the B-L limit), but not on the $\pi \rho V$ or $\pi \omega V$ correlator.
Our philosophy is that we take the first multiplets to resemble the corresponding ground-state physical particles but the second multiplets to mimic the effect of the tower of infinitely many excited states.
Since we study with $PVV$ an order parameter of chiral symmetry breaking~\cite{Knecht:2001xc} one expects that the second multiplets are close to the physical states that are the first excitations on top of the ground states.
However, the weighted average of the whole tower of states might shift the effective mass higher up.
In the present work we explore the uncertainty of the THS approximation by changing the masses of the second multiplets from the first to the second physical excitations.
Concerning the $\pi VV$ correlator we expect to obtain reasonable results because the pion might not resolve too many details of the intermediate-energy region.
Thus replacing the tower of excited resonances by the respective lowest excitation and demanding high-energy constraints for the $\pi VV$ correlator might be good enough.
In contrast, a $\pi \rho V$ or $\pi \omega V$ correlator resolves more from the intermediate-energy region because the vector meson induces a larger mass scale.
(The same would apply to a $\rho PV$ correlator.)
Therefore to satisfy high-energy constraints in this case we expect that one would need a more detailed modeling than just having THS, i.e.\ one excitation (plus the ground state) in each channel.
It might be worth to explore a three-hadron saturation scenario, but this is beyond the scope of the present work.
Therefore we demand constraints on the $PVV$ and $\pi VV$ correlators, but disregard constraints for three-point correlators of one pseudoscalar meson, one vector meson and one vector quark current.
For the same reason we only consider leading, but not subleading high-energy/OPE constraints.

From a formal point of view our approach is close to the successive Pad\'e approximations as utilized, e.g., in \cite{Masjuan:2015lca}.
Our correlators are also approximated by rational functions.
In our approach, however, the poles of the correlators are related to physical states (in the large-$N_c$ limit).
In contrast, in the Pad\'e framework of \cite{Masjuan:2015lca} one determines the rational functions by fits to data.
It is not the purpose of this Pad\'e approach to look for the poles of the obtained rational functions.
In fact, there are no physical restrictions from outside that would make sure that these poles correspond in any way to physical hadrons.
But this is what S-matrix theory suggests (in the large-$N_c$ limit): Anything beyond polynomials should be caused by unitarity, analyticity, crossing symmetry and physical states.

Our work is complementary to the dispersion theoretical approach of~\cite{Schneider:2012ez,Hoferichter:2014vra}; 
see also \cite{Danilkin:2014cra}.
While we cannot expect to reach the accuracy of a dispersive approach concerning low-energy quantities, our framework has the advantage to provide a smooth and physical connection between the low- and high-energy region and between the quark based 
and hadron based correlators.
In practice, dispersive calculations are based on an excellent account of the low-energy region (up to about 1 GeV) and a high-energy completion, i.e.\ a matching to the high-energy behavior deduced from S-matrix theory, QCD or QCD related approaches; see the discussion in~\cite{Hoferichter:2014vra} concerning the $\pi VV$ correlator.
As already discussed, for our doubly virtual $\pi VV$ correlator there are regions in the virtualities where the asymptotic regime is reached only at very high energies.
In practice this might imply that a naive matching of a dispersive calculation to the asymptotic regime might miss part of the physics present at intermediate energies.
Clearly it is worth to explore the interplay of a dispersive calculation with THS in the future.

The rest of the paper is structured in the following way: In Section~\ref{sec:ff} we construct the $PVV$ and $\pi VV$ correlators subject to high- and low-energy constraints.
The results for the $\pi VV$ correlator are compared to data and to other approaches (LMD and VMD).
It is studied how the shape of the correlator changes when varying the remaining free parameter $\kappa$ in a reasonable range.
In addition, a model uncertainty is estimated by varying the mass of the second vector-meson multiplet between the {\em physical} masses of the first and second excitation.

In Section~\ref{sec:pheno} the $\pi \omega V$ correlator is constructed.
The parameter $\kappa$ is determined from a fit to the $\omega$-$\pi$ transition form factor.
The widths or branching ratios, respectively, for the corresponding decays $\omega \to \pi^0 \gamma$, $\pi^0 e^+ e^-$, and $\pi^0 \mu^+ \mu^-$ are determined for THS (and also for LMD and VMD).
Actually the first decay does not depend on $\kappa$.
We find good agreement between THS and the experimental results.

We address the rare pion decay to electron and positron in Section~\ref{sec:piee}.
Including radiative corrections the branching ratio of this process is calculated and we compare again the 
THS result to other approaches.
For the THS case this branching ratio is sensitive to $\kappa$.
Direct comparison to the experimental value from KTeV \cite{Abouzaid:2006kk} seems to suggest that a discrepancy persists on the level of 2\,$\sigma$.

In Section~\ref{sec:pieegam} we study the properties of the singly virtual pion transition form factor in the low-energy region, such as slope and curvature.
Note that these quantities do not depend on the parameter $\kappa$.
For the Dalitz decay $\pi^0 \to e^+e^-\gamma$ we calculate the decay width taking into account next-to-leading order (NLO) radiative corrections.
These are evaluated along the lines of \cite{Husek:2015sma}.
Again we compare THS to other approaches.
With the full set of radiative corrections at hand, we take a fresh look on the KTeV result.
Considering some radiative corrections that were not accounted for in the analysis suggests that the previously stated discrepancy might be even reduced to 1.5\,$\sigma$.
The main message here, however, is that the radiative corrections are now theoretically under control \cite{Vasko:2011pi,Husek:2014tna,Husek:2015sma} and can be used in future data analyses.

In Section~\ref{sec:outlook} we provide an outlook how THS can be further utilized and how the scheme can be extended.
Finally, an appendix is added to provide the pseudoscalar form factors in terms of the loop integrals required for the rare pion decay.

Comparing in detail various approaches throughout the present work reveals that the VMD form factor proves to be phenomenologically very successful in most applications, in spite of the facts that VMD is so simple and partially possesses an improper high-energy behavior.
However, the THS form factor, which satisfies all the considered constraints, works very well, too.
Still, the mean value of the KTeV result remains a challenge for all approaches, even though the discrepancy has been reduced by a considerable level.
We provide a brief discussion on this point in the corresponding sections.


\section{THS approach to the \texorpdfstring{$\pi VV$}{piVV} correlator}
\label{sec:ff}

Following \cite{Knecht:2001xc,Kampf:2011ty} we introduce the $PVV$ correlator $\Pi(r^2;p^2,q^2)$ by 
\begin{equation}
\begin{split}
&d^{abc}\epsilon_{\mu\nu\alpha\beta }p^{\alpha}q^{\beta}\Pi(r^2;p^2,q^2)\\
&\equiv\int\text{d}^4x\,\text{d}^4y\,e^{ip\cdot x+iq\cdot y}\langle 0|T[P^a(0)V_\mu^b(x)V_\nu^c(y)]|0\rangle
\end{split}
\end{equation}
with $r=p+q$.
The vector and pseudoscalar current, respectively, are defined by%
\footnote{Our convention is $\gamma_5 = i \gamma^0\gamma^1\gamma^2\gamma^3$.}%
\begin{equation}
V_\mu^a(x)
\equiv\bar q(x)\gamma_\mu T^a q(x)\,,\quad
P^a(x)
\equiv\bar q(x)i\gamma_5 T^a q(x)\,.
\label{VPcur}
\end{equation}
In the above formulae we have used 
\begin{equation}
\text{Tr}\big[T^a,T^b\big]
=\frac12\delta^{ab}\,,\quad
d^{abc}
\equiv2\,\text{Tr}\big[\{T^a,T^b\}T^c\big]  \,.
\label{eq:normdef}
\end{equation}
For $a=1,\ldots,8$ we have $T^a\equiv{\lambda^a}/{2}$ where $\lambda^a$ denote the Gell-Mann matrices in flavor space.
Since we utilize the large-$N_c$ limit we have to deal with flavor nonets, i.e.\ with U(3) instead of SU(3).
The formulae (\ref{eq:normdef}) provide a natural extension to $a=0$.

Working in the chiral limit%
\footnote{The chiral limit is only used for the construction of the correlator.
Once the form factor is settled, we take the physical pion mass for all the kinematics used for comparing the predictions to experimental data.}
and considering two meson multiplets in the vector and two in the pseudoscalar channel, we propose a correlator of the form
\begin{equation}
\begin{split}
&\Pi^\text{THS}(r^2;p^2,q^2)=\frac{B_0 F^2}{r^2(r^2-M_P^2)}\\
&\times\frac{P(r^2;p^2,q^2)}{(p^2- M_{V_1}^2)(p^2- M_{V_2}^2)(q^2- M_{V_1}^2)(q^2- M_{V_2}^2)}\,.
\end{split}
\label{eq:corr}
\end{equation}
Here $P(r^2;p^2,q^2)$ is the most general polynomial in its arguments that is symmetric in the second and the third argument.
For our purpose, a polynomial of order four is sufficient inasmuch as higher powers of momenta in the numerator are not allowed due to the desired high-energy behavior that we will specify below.
Thus we start with a term containing 22 monomials and associated free parameters.
Schematically this looks as follows:
\begin{equation}
\begin{split}
P(r^2&;p^2,q^2)
=c_0p^2q^2+c_1[(p^2)^4+(q^2)^4]\\
&+c_2[(p^2)^3q^2+(q^2)^3p^2]+c_3(r^2)^2p^2q^2+\,\dots \,.
\end{split}
\label{eq:poly}
\end{equation}

The quantities $F$ and $B_0$ in (\ref{eq:corr}) denote the usual low-energy constants of chiral perturbation theory \cite{Gasser:1983yg,Gasser:1984gg}.
We will specify them further when needed.
$M_P$ denotes the mass of the second multiplet, i.e.\ first excitation, in the pseudoscalar channel.
The first multiplet (ground state) is, of course, the massless multiplet of Goldstone bosons.

The masses of the lowest two vector-meson multiplets are denoted by $M_{V_1}$ and $M_{V_2}$.
As already spelled out in Section~\ref{sec:intro} we use for $M_{V_1}$ the mass of the ground-state vector-meson multiplet (in the chiral limit).
We assume that this mass is approximately given by the mass of the $\omega$ or $\rho$ meson.
For $M_{V_2}$ it is suggestive to use the physical mass of the first excitation in the vector channel.
However, we do not use a fixed mass here, but study the impact of a variation of $M_{V_2}$ on our results.
In this way we explore the uncertainty caused by the higher-lying excitations that are neglected in THS.
In practice we vary $M_{V_2}$ in the range
\begin{equation}
M_{V_2} \in [1400,1740]\,\text{MeV,}
\label{eq:rangev2}
\end{equation}
which is the interval between the masses of the first and the second physical excitation \cite{PDG}.

Finally we note that the same logic applies to the mass $M_P$ of the second pseudoscalar multiplet.
For the $PVV$ correlator one should study the impact of a variation of $M_P$.
As we will see below, however, this mass does not show up in the final expression for the $\pi VV$ correlator.

We now demand that the ansatz (\ref{eq:corr}) satisfies all the relevant high- as well as low-energy constraints in order to minimize the number of free parameters introduced in (\ref{eq:poly}).
Starting with the general correlator (\ref{eq:corr}) we apply the following leading-order OPE constraints~\cite{Knecht:2001xc}: 
\begin{equation}
\begin{split}
\Pi\big((\lambda r)^2&;(\lambda p)^2,(\lambda q)^2\big)\\
&=\frac12{B_0F^2}\frac{1}{\lambda^4}\frac{r^2+p^2+q^2}{r^2p^2q^2}
+\mathcal{O}\bigg(\frac 1{\lambda^6}\bigg)\,,
\end{split}
\label{eq:OPE1}
\end{equation}
\begin{equation}
\Pi\big(r^2;(\lambda p)^2,(r-\lambda p)^2\big)
=B_0F^2\frac{1}{\lambda^2}\frac{1}{r^2p^2}
+\mathcal{O}\bigg(\frac 1{\lambda^3}\bigg)  \,.
\label{eq:OPE2}
\end{equation}
It turns out that the third OPE constraint \cite{Knecht:2001xc}
\begin{equation}
\Pi\big((q+\lambda p)^2;(\lambda p)^2,q^2\big)
=\frac{1}{\lambda^2}\frac{1}{p^2}f(q^2)
+\mathcal{O}\bigg(\frac 1{\lambda^3}\bigg)\,,
\label{eq:OPE3}
\end{equation}
is automatically fulfilled by our ansatz.
Here $f$ denotes a function (actually a two-point correlator \cite{Knecht:2001xc}) that depends only on $q^2$ and not on the other kinematic variables $p^2$ and $p\cdot q$.

Next, we define the $\pi VV$ correlator:
\begin{equation}
\mathcal{F}_{\pi VV}(p^2,q^2)
\equiv \frac1{\mathcal{Z}_{\pi}}\lim_{r^2\to0}r^2\Pi(r^2;p^2,q^2)\,,
\label{eq:FFp2q2}
\end{equation}
where
\begin{equation}
\mathcal{Z}_{\pi}
\equiv \frac i2\langle 0|(\bar u\gamma_5u-\bar d\gamma_5d)|\pi^0\rangle
\end{equation}
denotes the overlap between the pion field and the pseudoscalar quark current.
With the usual conventions from chiral perturbation theory one obtains $\mathcal{Z}_{\pi}=B_0F$~\cite{Gasser:1983yg,Gasser:1984gg}.
For large $\lambda$ one then finds
\begin{equation}
\begin{split}
&\mathcal{F}_{\pi VV}\big((\lambda p)^2,(\lambda p)^2\big)\\
&\simeq \frac1{\mathcal{Z}_{\pi}}\lim_{r^2\to0}r^2\Pi\big(r^2;(\lambda p)^2,(r-\lambda p)^2\big)\\
&\stackrel{(\ref{eq:OPE2})}{=}\frac{1}{B_0F}B_0F^2\frac{1}{\lambda^2p^2}
+\mathcal{O}\bigg(\frac 1{\lambda^3}\bigg)\,,
\end{split}
\end{equation}
which means that in agreement with~\cite{Lepage:1979zb,Lepage:1980fj,Knecht:2001xc} we have
\begin{equation}
\mathcal{F}_{\pi VV}(q^2,q^2)
\to \frac{F}{q^2}\,,\;q^2\to-\infty\,.
\label{eq:F/q2}
\end{equation}
Hence, this condition is satisfied automatically on account of the OPE constraints.
Actually, for the quantity $\mathcal{F}_{\pi VV}(q^2,q^2)$ also the subleading order in the high-energy expansion is known in terms of a quark-gluon condensate \cite{Novikov:1983jt}; see also the corresponding discussion in \cite{Jegerlehner:2009ry}.
However, we refrain from incorporating this as an additional constraint.
The reason is the same as to why we do not use constraints from transition form factors of hadronic resonance states.
One might become too sensitive to the details of the intermediate mass region where we use one hadronic state to describe effectively the whole infinite tower of large-$N_c$ excited states.
If one used three instead of two hadronic states per channel, the subleading high-energy constraints might provide a viable input to pin down the growing number of resonance parameters.
But this is clearly beyond the scope of the present work.

Instead of involving subleading orders in the high-energy expansion we apply the B-L constraint~\cite{Brodsky:1981rp}
\begin{equation}
\frac{\mathcal{F}_{\pi VV}(0,q^2)}{\mathcal{F}_{\pi VV}(0,0)}
\to-\frac{24\pi^2F^2}{N_c}\frac1{q^2}\,,\;{q^2\to-\infty}\,.
\label{eq:BL}
\end{equation}
We define the pion transition form factor as
\begin{equation}
\mathcal{F}_{\pi^0\gamma^*\gamma^*}(p^2,q^2)
=\frac23\mathcal{F}_{\pi VV}(p^2,q^2)
\label{eq:defpiTFF}
\end{equation}
and match at the photon point to the chiral anomaly, i.e.\ to the Wess--Zumino--Witten term~\cite{Wess:1971yu,Witten:1983tw},
\begin{equation}
\mathcal{F}_{\pi VV}(0,0)
=\frac{3}{2}\mathcal{F}_{\pi^0\gamma^*\gamma^*}(0,0)
=-\frac{N_c}{8\pi^2F}\,.
\label{eq:WZW}
\end{equation}

Together the constraints (\ref{eq:OPE1}), (\ref{eq:OPE2}), (\ref{eq:BL}) and (\ref{eq:WZW}) provide us with a $\pi VV$ correlator that appears to have only one free dimensionless parameter $\kappa$:
\begin{equation}
\begin{split}
&\mathcal{F}_{\pi VV}^\text{THS}(p^2,q^2)
=-\frac{N_c}{8\pi^2F}\\
&\times\frac{ M_{V_1}^4 M_{V_2}^4}{(p^2- M_{V_1}^2)(p^2- M_{V_2}^2)(q^2- M_{V_1}^2)(q^2- M_{V_2}^2)}\\
&\times\bigg\{1+\frac{\kappa}{2N_c}\frac{p^2q^2}{(4\pi F)^4}
-\frac{4\pi^2F^2(p^2+q^2)}{N_c M_{V_1}^2 M_{V_2}^2}\bigg[6+\frac{p^2q^2}{ M_{V_1}^2 M_{V_2}^2}\bigg]\bigg\}\,.
\end{split}
\label{eq:FFfinal2}
\end{equation}
Note that our result is independent of the mass $M_P$ of the first pseudoscalar excitation.
This happens due to the fact that at the end of the day we could conveniently rescale the only free parameter left.
From the structure of the result (\ref{eq:FFfinal2}) it can be read off that $\kappa$ emerges from $c_0$ in (\ref{eq:poly}) on account of
\begin{equation}
c_0=\frac{\kappa M_P^2 M_{V_1}^4 M_{V_2}^4}{(4\pi F)^6}\,.
\end{equation}
We note in passing that $\kappa$ scales with $N_c^3$.

A comparison to the work of \cite{Knecht:2001xc} is in order here concerning the number of free parameters.
We shall compare our THS approach to LMD+V of \cite{Knecht:2001xc}, i.e.\ to the case of one pseudoscalar and two vector multiplets.
We recall that THS = LMD+V+P in the language of \cite{Knecht:2001xc}.
After applying the OPE constraints to THS we are left with 12 free parameters for the $PVV$ correlator.
This compares to 7 free parameters for the case of LMD+V.
The low-energy constraint (\ref{eq:WZW}) fixes always one more parameter.
Once we focus on the $\pi VV$ correlator we are left with 4 parameters for THS.
For LMD+V one has 3.
Demanding that the $\pi VV$ correlator of (\ref{eq:BL}) drops like $1/q^2$ limits the free parameters to 2 for both cases THS and LMD+V.
Demanding (\ref{eq:BL}) {\em quantitatively} and not just a scaling with $1/q^2$ yields one free parameter for THS as well as for LMD+V.
To summarize, after applying all constraints THS has more parameters for the $PVV$ correlator than LMD+V: 8 in THS versus 4 in LMD+V.
But concerning the $\pi VV$ correlator one ends up with one free parameter in both cases.

According to \cite{Dorokhov:2007bd} the $\pi VV$ correlator should satisfy the inequality
\begin{equation}
|\mathcal{F}_{\pi VV}(q^2,q^2)| < |\mathcal{F}_{\pi VV}(0,q^2)|\,,\;q^2<0  \,.
\label{eq:DC}
\end{equation}
It turns out that the THS expression (\ref{eq:FFfinal2}) satisfies (\ref{eq:DC}) for $-45\lesssim\kappa\lesssim30$.
We will see below that the values for $\kappa$ that we obtain from fitting to experimental data lie well within this range.

Before further constraining $\kappa$ from data it is illuminating to study the qualitative shape of the $\pi VV$ correlator when $\kappa$ is varied.
In addition, we compare our results to similar approaches from the literature, namely to the VMD correlator \cite{sakuraiVMD,Landsberg:1986fd,Knecht:2001xc,Kampf:2011ty}
\begin{equation}
\mathcal{F}_{\pi VV}^\text{VMD}(p^2,q^2)
=-\frac{N_c}{8\pi^2F}\frac{ M_{V_1}^4}{(p^2- M_{V_1}^2)(q^2- M_{V_1}^2)}  
\label{eq:FFVMD}
\end{equation}
and to the LMD expression \cite{Knecht:1999gb,Knecht:2001xc}
\begin{equation}
\begin{split}
\mathcal{F}_{\pi VV}^\text{LMD}(p^2,q^2)
=\mathcal{F}_{\pi VV}^\text{VMD}(p^2,q^2)
\bigg[1-\frac{4\pi^2F^2(p^2+q^2)}{N_c M_{V_1}^4}\bigg]\,.
\label{eq:FFLMD}
\end{split}
\end{equation}
See also \cite{Bijnens:1999jp} for a comparison of various correlators.

\begin{figure}[!tb]
\resizebox{\columnwidth}{!}{\includegraphics{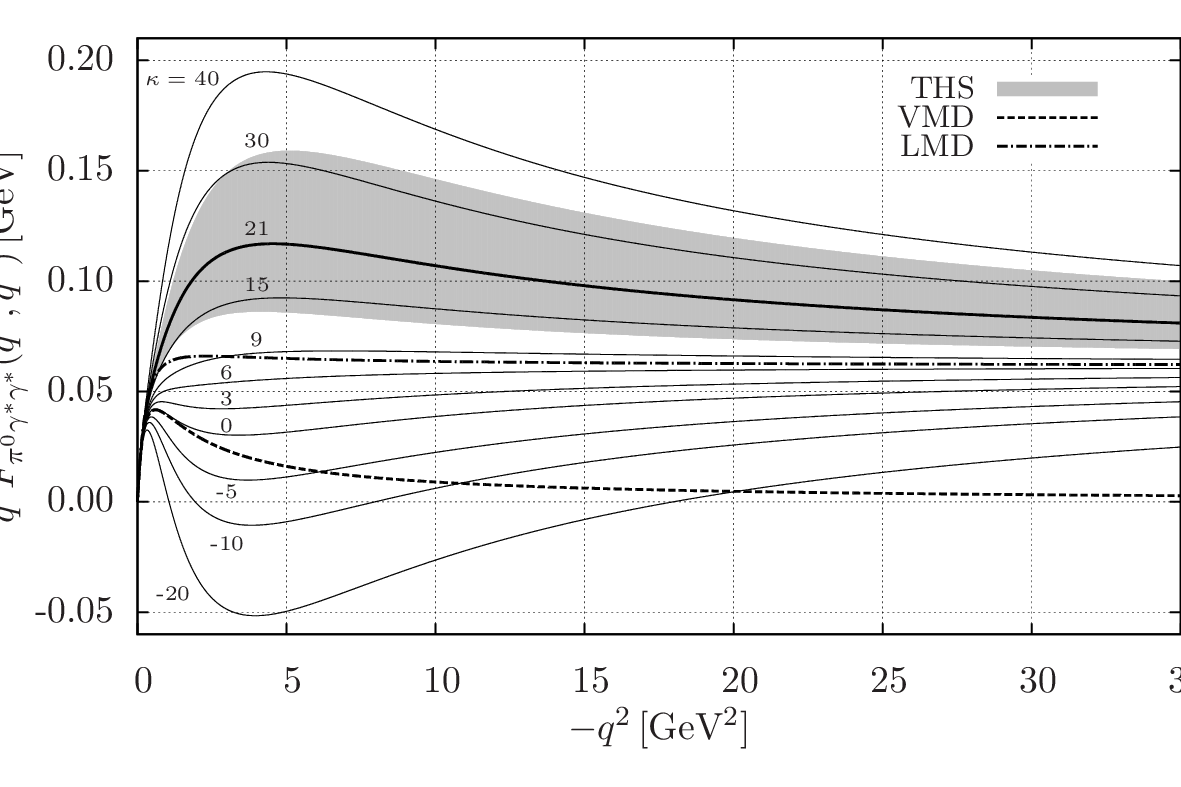}}
\resizebox{\columnwidth}{!}{\includegraphics{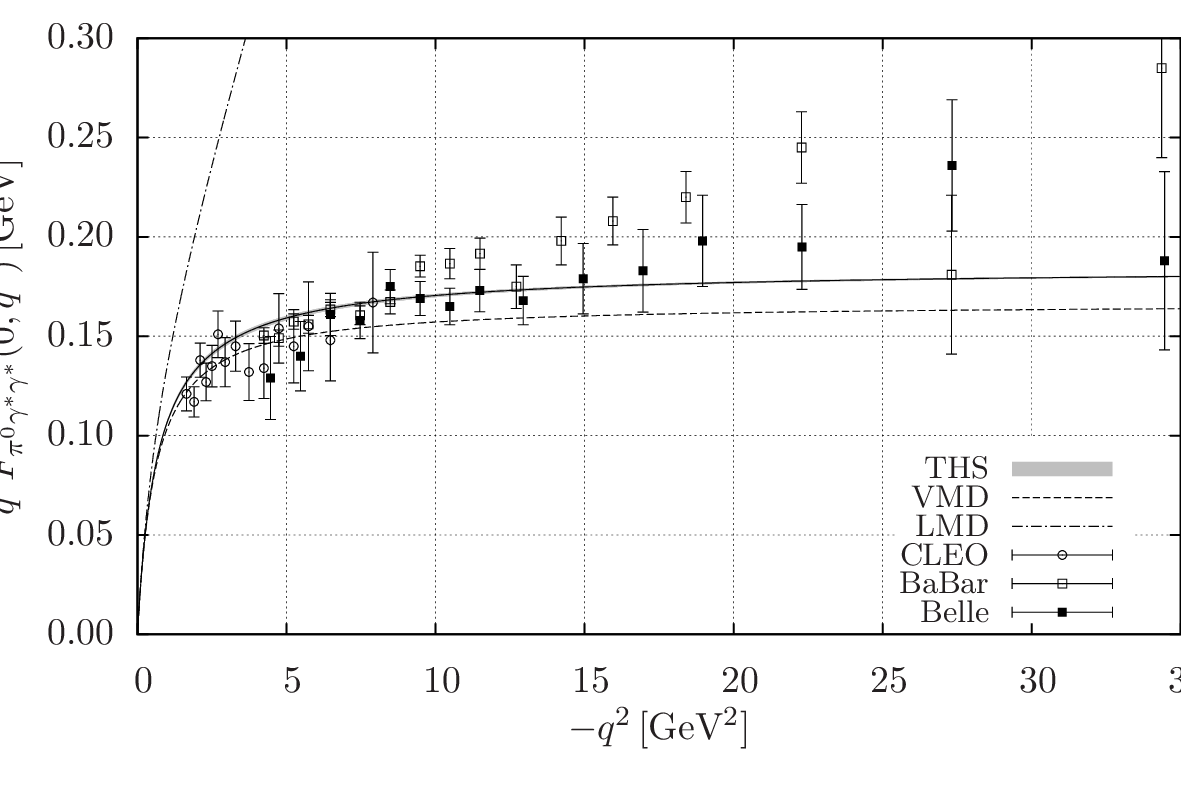}}
\caption{
Symmetric doubly virtual (first panel) and singly virtual (second panel) pion transition form factor as a function of the virtuality in the space-like region.
The gray bands constitute the final THS predictions.
See the main text for details how the uncertainties are determined.
In the second panel the gray band nearly collapses to one full line.
The thin full lines in the first panel show what happens if $\kappa$ is varied in a large range.
The corresponding curve that constitutes the center of the gray band has been made bold.
The results from VMD and LMD are given by the dashed and dash-dotted lines, respectively.
Data are taken from \cite{Gronberg:1997fj,Aubert:2009mc,Uehara:2012ag}.
}
\label{fig:FF}
\end{figure}

First of all, we note that the VMD result grossly violates (\ref{eq:F/q2}) while it is not far off from the B-L constraint (\ref{eq:BL}).
On the other hand, LMD satisfies (\ref{eq:F/q2}), but does not satisfy (\ref{eq:BL}).
By construction the THS correlator satisfies all the mentioned constraints, but it is interesting to see how fast or slow the asymptotic limits are reached.
This is shown in Fig.~\ref{fig:FF}.
To facilitate a comparison with data we show the pion transition form factor (times the virtuality $q^2$) instead of the $\pi VV$ correlator.
Note that the relation (\ref{eq:defpiTFF}) between the pion transition form factor and the $\pi VV$ correlator only amounts to a rescaling.

In the first panel of Fig.~\ref{fig:FF} we display the symmetric doubly virtual pion transition form factor.
This plot shows three different {\em types} of lines and a gray band.
We shall first discuss the lines: The dash-dotted LMD line approaches very quickly the asymptotic QCD result given by (\ref{eq:F/q2}).
The dashed VMD line falls stronger than what is required by QCD.
The full lines labeled by values for $\kappa$ show the THS result for a mass of $M_{V_2}$ taken in the middle of the interval (\ref{eq:rangev2}).
All the full lines will approach the LMD line at very large momenta.
One sees, however, that typically the THS lines reach this limit rather late.
In particular, for $\kappa\ge15$ the THS form factor peaks at rather high momentum values and with large magnitude in comparison to VMD and LMD.
It reaches the asymptotic limit only very slowly.
For the case $\kappa\simeq9$ we see that the associated line is very close to LMD, but peaks later and approaches the asymptotic limit from above.
For $\kappa\simeq6$ we get the limiting case where THS is monotonically growing.
Going further down with $\kappa$ a wiggle appears, which for $\kappa=0$ and low virtualities resembles very much the VMD behavior.
For negative $\kappa$ the THS results start to undershoot VMD at low virtualities.
Moving to even lower $\kappa$ values there is always a region where the THS result becomes negative.
Intuitively we find it hard to believe that the transition form factor of a ground-state hadron (the pion) would display so many wiggles as suggested by the full lines for small values of $\kappa$.
Indeed, the determination of $\kappa$, which will be carried out in the next section, reveals a $\kappa$ value of about 21.
Thus we find the qualitative situation that the THS result for the pion transition form factor (times the virtuality $q^2$) overshoots the asymptotic limit, peaks at rather large momenta and approaches the asymptotic limit rather slowly from above.

The gray bands in both panels of Fig.~\ref{fig:FF} constitute our final THS predictions for the pion transition form factor.
The results from the next section are anticipated where $\kappa$ is further constrained.
To obtain the gray bands all input for our THS pion transition form factor ($F$, $M_{V_1}$) is varied within the respective experimentally allowed regions, $\kappa = 21 \pm 3$ as described in the next section, and $M_{V_2}$ is varied within region (\ref{eq:rangev2}).
The whole gray band in the first panel shows the qualitative behavior described previously.
It also points to a significant quantitative uncertainty of our prediction for the symmetric doubly virtual pion transition form factor.
In other words, data on this form factor or in general on any doubly virtual pion transition form factor would be highly welcome to better constrain our approach.

In the second panel of Fig.~\ref{fig:FF} we display the singly virtual pion transition form factor.
This plot shows two lines, a very narrow gray band and data.
We first discuss the gray band, which nearly resembles a full line since its width is so small.
Obviously the narrow gray band of the second panel contrasts with the rather broad band of the first panel.
However, the singly virtual pion transition form factor is independent of $\kappa$ as can be easily deduced from (\ref{eq:FFfinal2}) when putting $p^2$ to zero.
The largest uncertainty comes from a variation of $M_{V_2}$ according to (\ref{eq:rangev2}).
Even this variation does not lead to very different results, which causes the curves to nearly collapse to one single line.
Thus THS has rather high predictive (or rather {\em postdictive}) power for the singly virtual pion transition form factor.
Another qualitative difference between the gray bands of the first and the second panel is the fact that the THS result for the singly virtual pion transition form factor reaches the asymptotic (B-L) limit rather early.
We regard this intriguing behavior of our correlator as one of the highlights of our work: The THS pion transition form factor shows an early (late) onset of the asymptotic behavior for the singly (symmetric doubly) virtual pion transition form factor.
It would be extremely interesting to see if this is supported by data in the future.

Finally, we compare the THS result for the singly virtual pion transition form factor to VMD and LMD and to data.
The LMD result for $\mathcal{F}_{\pi^0\gamma^*\gamma^*}(0,q^2)$ behaves as a constant for large $|q^2|$.
Thus, it cannot even qualitatively explain the data since, of course, it diverges after being multiplied by $q^2$.
This behavior can be seen in the second panel of Fig.~\ref{fig:FF} where the LMD result is given by the dash-dotted line.
The results for VMD (dashed line) and THS are fairly close.
CLEO data~\cite{Gronberg:1997fj} are only available for low virtualities ($-q^2<8$\,GeV$^2$) and prefer VMD to some extent though THS appears to be acceptable as well.
The situation turns around for higher virtualities.
THS describes the Belle data~\cite{Uehara:2012ag} better than VMD.
The early onset of the asymptotic behavior of the THS result for the singly virtual pion transition form factor makes THS essentially identical to the B-L limit.
Not surprisingly then, THS agrees well with the Belle data and is at odds with the BaBar data~\cite{Aubert:2009mc} that are not fully compatible with the Belle data.
We do not attempt to contribute to a clarification of the differences between Belle and BaBar.
In view of these complications we have decided to stick to the B-L constraint (\ref{eq:BL}) right from the start.


\section{Phenomenology of \texorpdfstring{$\omega$}{omega} decays}
\label{sec:pheno}

Our strategy now is to obtain the parameter $\kappa$ of the newly proposed $\pi VV$ correlator (\ref{eq:FFfinal2}) from the $\pi \omega V$ correlator.
Thus we turn to $\omega$ data for the reasons specified in Section~\ref{sec:intro}, namely the lack of doubly virtual data for the pion transition form factor and the fact that the $\omega$ meson is fairly long lived to resemble the situation of the large-$N_c$ approximation.

We introduce the overlap between an $\omega$ meson and the vector current $V$, i.e.
\begin{equation}
\mathcal{Z}_{\omega}\epsilon_\mu(\vec{p},\lambda_\omega)
\equiv \frac 12\langle 0|(\bar u\gamma_\mu u+\bar d\gamma_\mu d)|\omega(\vec{p},\lambda_\omega)\rangle\,,
\end{equation}
where we assume that the $\omega$ meson does not contain hidden strangeness, which is a fairly good approximation to the real world \cite{PDG}.
For later convenience we also define the $\gamma$-$\omega$ coupling strength
\begin{equation}
F_\omega
\equiv \frac{Z_\omega}{M_\omega}\,,
\end{equation}
the modulus of which agrees with $F_V$ as introduced in~\cite{Ecker:1988te}.

With this at hand we can obtain $F_{\omega}$ from the $\omega\to e^+e^-$ decay process.
A direct calculation from the Lorentz invariant matrix element
\begin{equation}
i\mathcal{M}_{\omega\to e^+e^-}
=\frac{ie\mathcal{Z}_\omega}{3}\bar{u}(\vec{q}_1)(-ie\gamma^\mu)v(\vec{q}_2)\frac{(-i)}{M_\omega^2}\epsilon_\mu(\vec{p},\lambda_\omega)\,,
\end{equation}
yields after averaging and summing over polarizations
\begin{equation}
F_{\omega}
=\bigg[\frac{4\pi\alpha^2}{27M_\omega}\frac1{\Gamma({\omega\to e^+e^-})}\beta_{e\omega}\bigg(1+\frac{2m_e^2}{M_\omega^2}\bigg)\bigg]^{-\frac12}\,,
\end{equation}
where $\beta_{e\omega}=\sqrt{1-{4m_e^2}/{M_\omega^2}}$ is the speed of the electron in the rest frame of the decaying $\omega$ meson.
Taking into account that $m_e^2\ll M_\omega^2$, we can write
\begin{equation}
F_{\omega}
\simeq
\frac{1}{e^2}\sqrt{{108\pi M_\omega{\Gamma({\omega\to e^+e^-})}}}\,.
\label{eq:FV1}
\end{equation}
Using the values $B({\omega\to e^+e^-})=(7.28\pm0.14)\times10^{-5}$ and $\Gamma(\omega)=(8.49\pm0.08)$\,MeV~\cite{PDG} we find $\Gamma({\omega\to e^+e^-})=(0.62\pm0.02)$\,keV.
Together with $M_\omega=(782.65\pm0.12)$\,MeV we obtain from (\ref{eq:FV1}) the following value for the coupling strength:
\begin{equation}
F_{\omega}
=(140\pm2)\,\text{MeV.}
\end{equation}

It is convenient to introduce the $\pi^0\omega V$ transition form factor by 
\begin{equation}
\mathcal{F}_{\pi^0\omega V}(q^2)
=\frac1{\mathcal{Z}_\omega}\lim_{p^2\to M_{V_1}^2}(p^2- M_{V_1}^2)\mathcal{F}_{\pi VV}(p^2,q^2)\,.
\label{eq:FFq2}
\end{equation}
This turns out to be the central quantity to determine $\kappa$ and to address the phenomenologically interesting decays $\omega\to\pi^0\gamma$ and $\omega\to\pi^0\ell^+\ell^-$, where $\ell$ denotes a lepton.

For the special case when $q^2=0$ we find (see also \cite{RuizFemenia:2003hm} for a similar expression for the case of one vector multiplet)
\begin{equation}
\mathcal{F}_{\pi^0\omega V}^\text{THS}(0)
=\frac1{\mathcal{Z}_\omega}\frac{N_c}{8\pi^2F}\frac{ M_{V_1}^2 M_{V_2}^2}{( M_{V_2}^2- M_{V_1}^2)}\bigg[1-\frac{24\pi^2F^2}{N_c M_{V_2}^2}\bigg]\,.
\end{equation}
Hence the decay width for $\omega\to\pi^0\gamma$ is independent of $\kappa$, which provides us with full predictive power for this particular decay and can therefore serve as a cross-check of our formalism.
We will come back to this decay after having determined $\kappa$ from the measured $\omega$-$\pi$ transition form factor.

\begin{figure}[t]
\resizebox{\columnwidth}{!}{\includegraphics{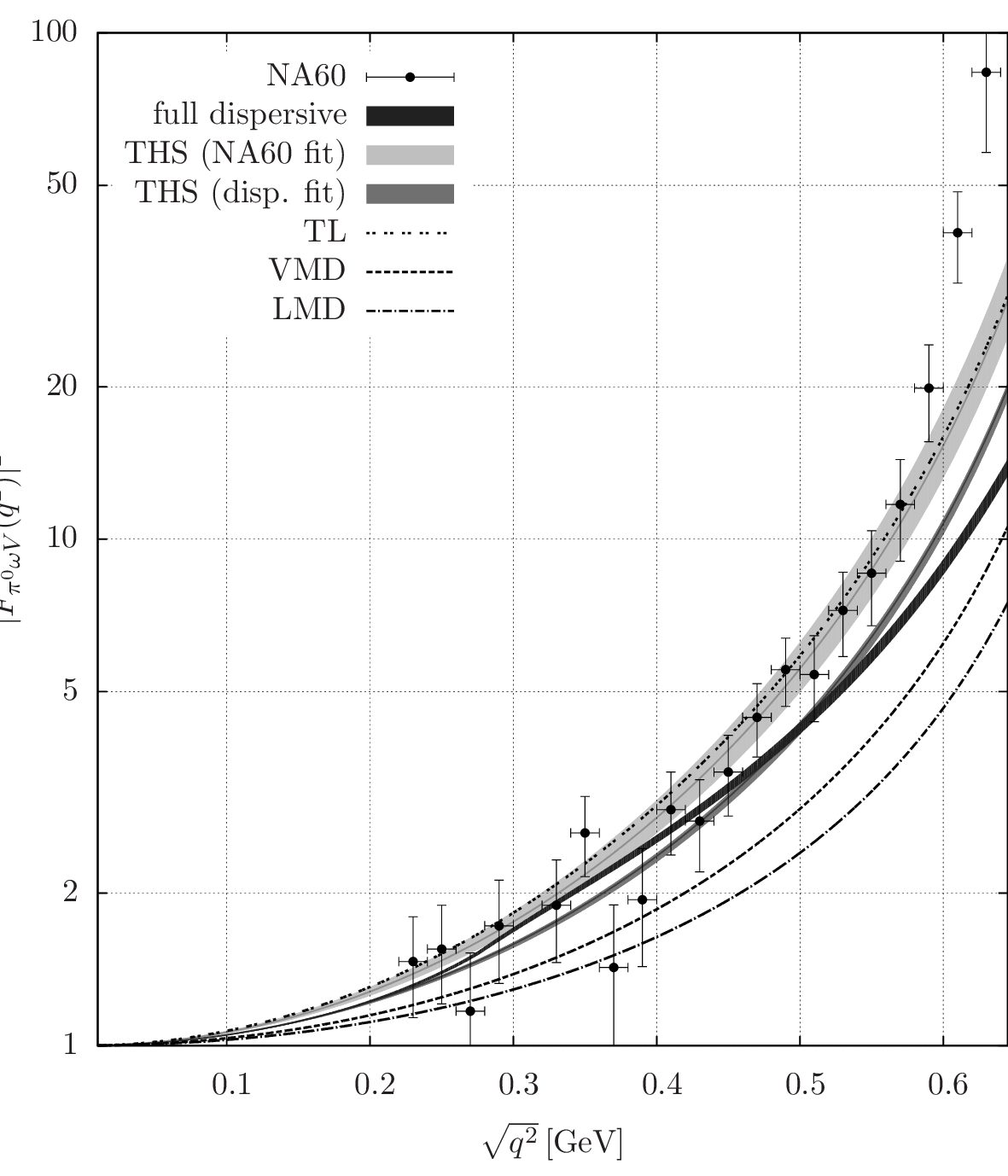}}
\caption{
The normalized $\omega$-$\pi$ transition form factor as a function of the (time-like) virtuality.
Various theoretical calculations are displayed together with NA60 data obtained from the decay $\omega \to \pi^0 \mu^+ \mu^-$ \cite{Arnaldi:2009aa}.
See the main text for more details.
}
\label{fig:FFomega}
\end{figure}

In fact, the parameter $\kappa$, which constitutes the only unknown parameter of our proposed $\pi VV$ correlator, can be obtained by a fit to the NA60 data \cite{Arnaldi:2009aa} for the normalized transition form factor
\begin{equation}
\mathcal{\hat F}_{\pi^0\omega V}(q^2)
\equiv\frac{\mathcal{F}_{\pi^0\omega V}(q^2)}{\mathcal{F}_{\pi^0\omega V}(0)}
\label{eq:ompinorm}
\end{equation}
in the low-energy time-like region of $q^2$ with the result
\begin{equation}
\kappa
=21\pm3  \,.
\label{eq:kappa}
\end{equation}
The result is displayed in Fig.~\ref{fig:FFomega} as a gray band.
The uncertainty comes mainly from the fitting procedure (and so from the error bars of the data) and from the variation of the second-multiplet mass $ M_{V_2}$ inside the considered region (\ref{eq:rangev2}).
Our fit to the NA60 data has a $\chi^2$ per degree of freedom of 1.5.

Apparently we obtain a rather satisfying fit to the NA60 data except for the last two or three data points at the largest values of the dimuon mass.
At present, none of the hadron-theory approaches to this $\omega$ transition form factor \cite{Terschluesen:2010ik,Ivashyn:2011hb,Schneider:2012ez,Chen:2013nna,Danilkin:2014cra} is able to understand these last data points.
Recently it has been suggested in \cite{Ananthanarayan:2014pta} using a dispersive calculation and high-energy constraints that these data points might be incompatible with QCD.
Clearly it would be highly desirable to obtain additional data for this $\omega$ transition form factor, in particular from experiments where the complete final state $\pi^0 \ell^+ \ell^-$ can be reconstructed and not only the dilepton; see \cite{Arnaldi:2009aa} for more details.

In the following we will stick to our result for $\kappa$ as given in (\ref{eq:kappa}) and obtained from a fit to the full range of NA60 data since it is the best that one can do in the present situation.
However, we briefly discuss two alternatives: If we performed an alternative fit to the NA60 data rejecting the last three data points, we would get $\kappa=19\pm2\,$, which is fairly compatible with our full fit.
The $\chi^2$ per degree of freedom would reduce to 0.8.
If we rejected the NA60 data altogether and regarded the dispersive calculation of \cite{Schneider:2012ez} as the ``truth'', we would find $\kappa=13.1\pm0.5$\,.
The result of this last fit is also shown in Fig.~\ref{fig:FFomega} as the dark gray band.
If we return to the first panel of Fig.~\ref{fig:FF} we observe that one obtains the same qualitative features with such values of $\kappa$.
Nevertheless, we will not use these results further on and will take into account only the result (\ref{eq:kappa}) of the all-data fit.

As for the pion transition form factor we shall compare our result to similar approaches from the literature.
The form factor from \cite{Terschluesen:2010ik}, 
\begin{equation}
\mathcal{\hat F}_{\pi^0\omega V}^\text{TL}(q^2)
=\frac{ M_{V_1}^2+q^2}{ M_{V_1}^2-q^2}   \,,
\label{eq:FFTL}
\end{equation} 
labeled by ``TL'' in Fig.~\ref{fig:FFomega}, lies within our uncertainty band in spite of the fact that the derivation is based on a very different approach.
The results for VMD and LMD can be obtained from (\ref{eq:FFVMD}) and (\ref{eq:FFLMD}), respectively, using (\ref{eq:FFq2}) and (\ref{eq:ompinorm}).
For instance, the VMD expression is
\begin{equation}
\mathcal{\hat F}_{\pi^0\omega V}^\text{VMD}(q^2)
=\frac{ M_{V_1}^2}{ M_{V_1}^2-q^2}\,.
\label{eq:wVMD}
\end{equation}
As can be seen in Fig.~\ref{fig:FFomega} the results for both VMD and LMD deviate significantly from the data and therefore from our approach.
They also deviate from the results of the dispersive calculation of \cite{Schneider:2012ez}, which is shown as the black band in Fig.~\ref{fig:FFomega}.
As already noted we have performed an alternative fit of our THS expression to the dispersive result --- the dark gray band in Fig.~\ref{fig:FFomega}.
It should be noted that our rational function cannot fit the cusp structure in the low-energy region $\sqrt{q^2}\in[0.3,0.4]$ that emerges in the dispersive calculation from a cross-channel inelasticity; see \cite{Schneider:2012ez} for a detailed discussion.
We repeat our statement that additional data on this transition form factor would be highly welcome.

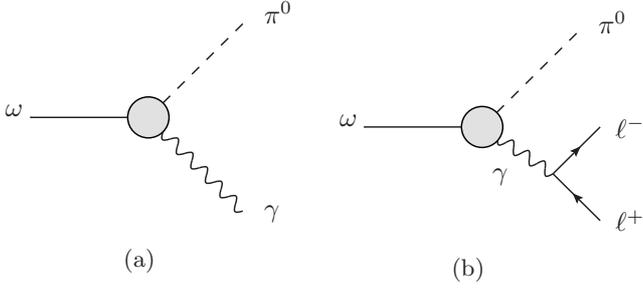
\begin{figure}[t]
\begin{subfigure}{0.45\columnwidth}
\setlength{\unitlength}{0.55pt}
\begin{picture}(208,152) (180,-230)
    \SetScale{0.55}
    \SetWidth{1.0}
    \SetColor{Black}
    \Line(208,-152)(288,-152)
    \Line[dash,dashsize=7](288,-152)(352,-88)
    \Text(192,-152)[lb]{\normalsize{\Black{$\omega$}}}
    \Text(368,-88)[lb]{\normalsize{\Black{$\pi^0$}}}
    \Text(368,-224)[lb]{\normalsize{\Black{$\gamma$}}}
    \Photon(352,-216)(288,-152){4}{6.5}
    \GOval(288,-152)(14,14)(0){0.882}
\end{picture}
\caption{}
\end{subfigure}
\begin{subfigure}{0.5\columnwidth}
\setlength{\unitlength}{0.55pt}
\begin{picture}(220,165) (165,-230)
    \SetScale{0.55}
    \SetWidth{1.0}
    \SetColor{Black}
    \Line(208,-152)(288,-152)
    \Line[dash,dashsize=7](288,-152)(352,-88)
    \Text(192,-152)[lb]{\normalsize{\Black{$\omega$}}}
    \Text(368,-88)[lb]{\normalsize{\Black{$\pi^0$}}}
    \Text(296,-192)[lb]{\normalsize{\Black{$\gamma$}}}
    \Photon(336,-184)(288,-152){4}{4.5}
    \Line[arrow,arrowpos=0.5,arrowlength=5,arrowwidth=2,arrowinset=0.2](336,-184)(368,-152)
    \Line[arrow,arrowpos=0.5,arrowlength=5,arrowwidth=2,arrowinset=0.2](368,-216)(336,-184)
    \Text(380,-160)[lb]{\normalsize{\Black{$\ell^-$}}}
    \Text(380,-224)[lb]{\normalsize{\Black{$\ell^+$}}}
    \GOval(288,-152)(14,14)(0){0.882}
\end{picture}
\caption{}
\end{subfigure}
\caption{
Feynman diagrams for the $\omega\to\pi^0\gamma$ and $\omega\to\pi^0\ell^+\ell^-$ decays at leading order in QED.
The shaded blob corresponds to the $\mathcal{F}_{\pi^0\omega\gamma^{(*)}}(q^2)$ form factor and $\ell$ denotes $e$ or $\mu$.
}
\label{fig:omega}
\end{figure}

To further explore the validity of the THS scheme we study selected decay channels of the $\omega$ meson.
It appears to be convenient to introduce the matrix element
\begin{equation}
\begin{split}
\mathcal{M}_{\omega\pi^0}^{\mu\nu}(p,q)
&=\frac{e}{\mathcal{Z}_\omega \mathcal{Z}_{\pi}}\bigg(\frac23+\frac13\bigg)\epsilon^{\mu\nu\alpha\beta}p_\alpha q_\beta\\
&\times\underset{p^2\to M_{V_1}^2}{\lim_{r^2\to0}}(p^2- M_{V_1}^2)r^2\Pi(r^2;p^2,q^2)\,.
\label{eq:Mpq1}
\end{split}
\end{equation}
Using previous definitions (\ref{eq:FFp2q2}) and (\ref{eq:FFq2}) this can be reduced to 
\begin{equation}
\mathcal{M}_{\omega\pi^0}^{\mu\nu}(p,q)
=e\mathcal{F}_{\pi^0\omega V}(q^2)\epsilon^{\mu\nu\alpha\beta}p_\alpha q_\beta\,.
\end{equation}
In (\ref{eq:Mpq1}) the factors 2/3 and 1/3 emerge from the respective overlap of the $\omega$ meson with the singlet and the 8th component of the octet current, i.e.\ the factors come from (\ref{eq:normdef}).
These factors actually sum to unity, which is not surprising from a flavor SU(2) point of view where the $\omega$ is a pure singlet.

We start with the prediction of the decay width of the $\omega\to\pi^0\gamma$ process, which is depicted in Fig.~\ref{fig:omega}a.
The matrix element
\begin{equation}
\mathcal{M}_{\omega\to\pi^0\gamma}
=\mathcal{M}_{\omega\pi^0}^{\mu\nu}(p,q)\epsilon_\mu(\vec{p},\lambda_\omega)\epsilon_\nu^*(\vec{q},\lambda_\gamma)\Big|_{q^2=0}
\end{equation}
does not depend on $\kappa$ as already mentioned.
This provides us with a pure prediction.
After simple manipulations we find 
\begin{equation}
\begin{split}
\Gamma(\omega\to\pi^0\gamma)
&=\frac{\alpha M_\omega^3}{24}|\mathcal{F}_{\pi^0\omega V}(0)|^2\bigg(1-\frac{M_\pi^2}{M_\omega^2}\bigg)^3\\
&\simeq
\frac{\pi\alpha^3}{162}\frac{\mathcal{Z}_\omega^2|\mathcal{F}_{\pi^0\omega V}(0)|^2}{\Gamma({\omega\to e^+e^-})}\bigg(1-\frac{M_\pi^2}{M_\omega^2}\bigg)^3\,,
\end{split}
\label{eq:alwaysonemorelabelwhichihavetoadd}
\end{equation}
where we have used (\ref{eq:FV1}) to obtain the last expression.
Taking $N_c=3$, $F=(92.22\pm0.06)$\,MeV~\cite{Kampf:2012pa}, $M_\pi=134.98$\,MeV, $ M_{V_1} = M_\rho=(775.26\pm0.25)$\,MeV~\cite{PDG} and $ M_{V_2}$ in the range (\ref{eq:rangev2}) we obtain the values listed in Tab.~\ref{tab:Gwpi0g}.
\begin{table}[!ht]
\begin{center}
{\scriptsize
\begin{tabular}{c | c c c c }
\toprule
 & VMD & LMD & THS & experiment\\
\midrule
$\Gamma(\omega\to\pi^0\gamma)$\,[MeV] & 0.68 & 0.45 & $0.63\pm0.04$ & $0.70\pm0.03$\\
\bottomrule
\end{tabular}}
\end{center}
\caption{
Decay width of the process $\omega\to\pi^0\gamma$ as obtained from different approaches and from data~\cite{PDG}.
We estimate an uncertainty only for our THS case.
}
\label{tab:Gwpi0g}
\end{table}
We see that THS is within its uncertainties compatible with the experimental data.
In this case VMD proves to be phenomenologically successful, too.
The value of the LMD approach seems to be off even taking into account the 40\% rule-of-thumb uncertainty used in~\cite{Knecht:1999gb}.

Moving on to the $\omega\to\pi^0\ell^+\ell^-$ decay, which is depicted in Fig.~\ref{fig:omega}b, the matrix element can be written in the form
\begin{equation}
\mathcal{M}_{\omega\to\pi^0\ell^+\ell^-}
=\mathcal{M}_{\omega\pi^0}^{\mu\nu}(p,q)\epsilon_\mu(\vec{p},\lambda_\omega)\frac{(-i)}{q^2}\bar{u}(\vec{q}_1)(-ie\gamma_\nu)v(\vec{q}_2)\,.
\end{equation}
The decay width can then be expressed as a form-factor-dependent integral over the dilepton invariant mass, 
\begin{equation}
\begin{split}
&\Gamma(\omega\to\pi^0\ell^+\ell^-)
=\frac{\alpha^2}{72\pi M_\omega^3}
\hspace{-3mm}
\int\limits_{4m_\ell^2}^{(M_\omega-M_\pi)^2}
\hspace{-2mm}
\;
\frac{|\mathcal{F}_{\pi^0\omega V}(q^2)|^2}{q^2}\\
&\times\sqrt{1-\frac{4m_\ell^2}{q^2}}
\bigg(1+\frac{2m_\ell^2}{q^2}\bigg)
\lambda^{\frac32}\big(M_\omega^2,M_\pi^2,q^2\big)\,
{\text{d}q^2}\,,
\end{split}
\end{equation}
where $\lambda$ denotes the K\"all\'en triangle function defined as
\begin{equation}
\lambda(a,b,c)
\equiv a^2+b^2+c^2-2ab-2ac-2bc\,.
\end{equation}
For THS the result for this decay width depends on the parameter $\kappa$.
We use the fitted value (\ref{eq:kappa}) and for the other input quantities the ranges specified after (\ref{eq:alwaysonemorelabelwhichihavetoadd}).

It is common practice to normalize decay widths that involve dileptons to the corresponding decay widths involving photons \cite{Landsberg:1986fd}.
This leads to the branching ratios listed in Tab.~\ref{tab:Bwpi0ll}.
\begin{table}[!ht]
\begin{center}
{\scriptsize
\begin{tabular}{c | c c c c }
\toprule
 & VMD & LMD & THS & experiment\\
\midrule
$\frac{B(\omega\to\pi^0e^+e^-)}{B(\omega\to\pi^0\gamma)}\times 10^3$ & $9.1$ & $8.9$ & $9.6\pm0.1$ & $9.3\pm1.0$ \\
$\frac{B(\omega\to\pi^0\mu^+\mu^-)}{B(\omega\to\pi^0\gamma)}\times 10^3$ & $0.91$ & $0.82$ & $1.33\pm0.10$ & $1.6\pm0.5$ \\
\bottomrule
\end{tabular}}
\end{center}
\caption{
Branching ratios of the decays $\omega\to\pi^0e^+e^-$ and $\omega\to\pi^0\mu^+\mu^-$ normalized to the branching ratio of the decay $\omega\to\pi^0\gamma$ as obtained from different approaches and from data~\cite{PDG}.
We estimate an uncertainty only for our THS case.}
\label{tab:Bwpi0ll}
\end{table}
Obviously, for the electron case all the results of the considered approaches lie within the experimental uncertainty.
For the muon case, however, only the THS result explains the experimental value.
This should not come as a surprise given that in Fig.~\ref{fig:FFomega} the LMD and VMD curves lie much lower than the THS band.


\section{The process \texorpdfstring{$\pi^0\to e^+e^-$}{pi0 --> e+e-}}
\label{sec:piee}

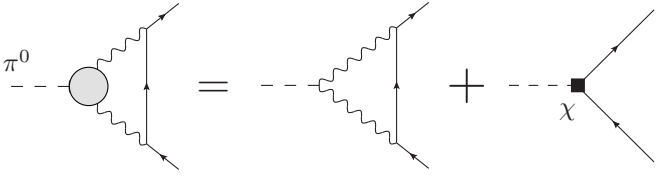
\begin{figure}[h]
\setlength{\unitlength}{0.40pt}
\begin{picture}(630,150) (161,-170)
    \SetScale{0.40}
    \SetWidth{1.0}
    \SetColor{Black}
    \Text(580,-100)[lb]{\huge{\Black{$+$}}}
    \Text(346,-95)[lb]{\LARGE{\Black{$=$}}}
    \Line[dash,dashsize=10](636,-83)(699,-83)
    \Line[arrow,arrowpos=0.5,arrowlength=5,arrowwidth=2,arrowinset=0.2](699,-84)(771,-12)
    \Line[arrow,arrowpos=0.5,arrowlength=5,arrowwidth=2,arrowinset=0.2](771,-155)(699,-83)
    \CBox(694,-89)(706,-77){Black}{Black}
    \Text(685,-116)[lb]{\normalsize{\Black{$\chi$}}}
    \Line[dash,dashsize=10](405,-83)(459,-83)
    \Photon(459,-83)(531,-29){4}{5.5}
    \Line[arrow,arrowpos=0.5,arrowlength=5,arrowwidth=2,arrowinset=0.2](561,-161)(531,-137)
    \Line[arrow,arrowpos=0.5,arrowlength=5,arrowwidth=2,arrowinset=0.2](531,-137)(531,-29)
    \Line[arrow,arrowpos=0.5,arrowlength=5,arrowwidth=2,arrowinset=0.2](531,-29)(561,-5)
    \Photon(531,-137)(459,-83){4}{5.5}
    \Line[dash,dashsize=10](172,-84)(226,-84)
    \Photon(226,-84)(298,-30){4}{5.5}
    \Line[arrow,arrowpos=0.5,arrowlength=5,arrowwidth=2,arrowinset=0.2](328,-162)(298,-138)
    \Line[arrow,arrowpos=0.5,arrowlength=5,arrowwidth=2,arrowinset=0.2](298,-138)(298,-30)
    \Line[arrow,arrowpos=0.5,arrowlength=5,arrowwidth=2,arrowinset=0.2](298,-30)(328,-6)
    \Photon(298,-138)(226,-84){4}{5.5}
    \GOval(244,-84)(18,18)(0){0.882}
    \Text(165,-70)[lb]{\normalsize{\Black{$\pi^0$}}}
\end{picture}
\caption{
Leading-order contribution to the $\pi^0\to e^+e^-$ process in the QED expansion and its representation in terms of the leading order in chiral perturbation theory.
The shaded blob corresponds to the doubly virtual pion transition form factor $\mathcal{F}_{\pi^0\gamma^*\gamma^*}(l^2,(r-l)^2)$.
}
\label{fig:piee}
\end{figure}

In the following two sections we turn our attention to the decays of the neutral pion.
Namely we will discuss the rare decay $\pi^0\to e^+e^-$, to which this section is devoted, and later on also the Dalitz decay $\pi^0\to e^+e^-\gamma$.

The decay $\pi^0\to e^+e^-$ and the radiative corrections connected to this process have been extensively studied in~\cite{Vasko:2011pi,Husek:2014tna}.
On the left-hand side of the graphical equation in Fig.~\ref{fig:piee}
the leading order of the QED expansion of the considered process is depicted.
The doubly virtual pion transition form factor $\mathcal{F}_{\pi^0\gamma^*\gamma^*}(l^2,(r-l)^2)$ is represented in the figure as the shaded blob.
Here $r$ denotes the pion momentum and $l$ a loop momentum.
This transition form factor plays here an essential role, since it serves as an effective ultraviolet (UV) cut-off due to its $1/l^2$ asymptotics governed by the OPE.
The loop integral over $\text{d}^4l$ is therefore convergent.

On account of Lorentz and parity symmetry the on-shell matrix element of the $\pi^0\to e^+e^-$ process can be written in terms of just one pseudoscalar form factor in the following form:
\begin{equation}
i\mathcal{M}_{\pi^0\to e^+e^-}
=P_{\pi^0\to e^+e^-}\left[\bar{u}(\vec{q}_1)\gamma_5 v(\vec{q}_2)\right]\,.
\label{eq:M1gI}
\end{equation}
Subsequently, the decay width reads
\begin{equation}
\Gamma(\pi^0\to e^+e^-)
=\frac{2M_\pi^2|P_{\pi^0\to e^+e^-}|^2}{16\pi M_\pi}\sqrt{1-\frac{4m_e^2}{M_\pi^2}}\,.
\end{equation}
Taking into account only the leading order (LO) in the QED expansion, i.e.\ the left-hand side of the graphical equation in Fig.~\ref{fig:piee}, we find (within the dimensional reduction scheme~\cite{Siegel:1979wq,Novotny:1994yx})
\begin{equation}
\begin{split}
P_{\pi^0\to e^+e^-}^\text{LO}
=\frac{ie^4 m_e}{M_\pi^2}
\int\frac{\text{d}^4l}{(2\pi)^4}
\frac{\mathcal{F}_{\pi^0\gamma^*\gamma^*}(p^2,q^2)\lambda(M_\pi^2,p^2,q^2)}{p^2q^2(l^2-m_e^2)}\,.
\end{split}
\label{eq:PLO}
\end{equation}
Here $p=l-q_1$ and $q=l+q_2$, where $q_1$ and $q_2$ are the lepton momenta.
Using dimensional regularization and Passarino--Veltman reduction~\cite{Passarino:1978jh} the explicit result of the loop integration in terms of scalar one-loop integrals is given in \ref{app:FF} for various form factors.

On the right-hand side of Fig.~\ref{fig:piee} we can see how the previously discussed expression is represented in leading order of chiral perturbation theory \cite{Savage:1992ac,Knecht:1999gb}.
Here, the constant 
\begin{equation}
\mathcal{F}_{\pi^0\gamma^*\gamma^*}(0,0)= -\frac{N_c}{12\pi^2F}
\label{eq:WZWagain}  
\end{equation}
emerging from the chiral anomaly is used instead of the full transition form factor; see (\ref{eq:WZW}).
This leads to a divergent integral \cite{Drell:1959nc} and it is thus clear that a counter-term is needed.
If the loop is renormalized at scale $\mu$, the finite part of such a counter-term is denoted by $\chi^\text{(r)}(\mu)$.
It corresponds to the high-energy behavior of the complete transition form factor.
Therefore, concerning the pion decay, we can characterize each transition form factor by the corresponding value of $\chi^\text{(r)}(\mu)$.

In fact, using (\ref{eq:WZWagain}) instead of the full form factor in (\ref{eq:PLO}) together with the counter-term Lagrangian from~\cite{Savage:1992ac,Knecht:1999gb} one obtains 
\begin{equation}
\begin{split}
&P_{\pi^0\to e^+e^-}^\text{LO}
=-2\alpha^2m_e \mathcal{F}_{\pi^0\gamma^*\gamma^*}(0,0)
\left\{-\frac 52+\frac 32\log\left(\frac{m_e^2}{\mu^2}\right)\right.\\
&+\chi^\text{(r)}(\mu)
+\frac1{2\beta_e}\left[\text{Li}_2(z)-\text{Li}_2\left(\frac 1z\right)+i\pi\log(-z)\right]\bigg\}\,.
\end{split}
\label{eq:PLOconst}
\end{equation}
In the above formula $\text{Li}_2$ is the dilogarithm, $\beta_e=\sqrt{1-{4m_e^2}/{M_\pi^2}}$, $z=-(1-\beta_e)/(1+\beta_e)$ and $\mu$ represents the scale at which the loop integral is effectively cut off; cf.~\cite{Knecht:1999gb}.
This effective approach with the constant form factor leading to formula (\ref{eq:PLOconst}) can be conveniently used to match with the calculation (\ref{eq:PLO}) and any full, i.e.\ momentum-dependent form factor.
Thus, for various approaches, the value of the associated effective parameter $\chi^\text{(r)}(\mu)$ can be extracted.
In other words, the left-hand side of (\ref{eq:PLOconst}) can be substituted by the expressions for $P_{\pi^0\to e^+e^-}^\text{LO}$ stated in \ref{app:FF} and $\chi^\text{(r)}(\mu)$ is subsequently determined.
Following common practice \cite{Gasser:1983yg,Ecker:1988te,Savage:1992ac,Knecht:1999gb} we have chosen $\mu = 770\,$MeV $\simeq M_\rho$ to set the renormalization scale.

An approximative way to get the low-energy constant $\chi^\text{(r)}(\mu)$ up to corrections proportional to $m_e/M_\pi$ and $M_\pi/\mu$ has been presented in \cite{Dorokhov:2007bd}.
In our notation this reads
\begin{equation}
\chi^\text{(r)}(\mu)
\simeq\frac54+\frac32
\int_0^\infty\hskip-2pt\text{d}t\,\log(t)\,\frac{\partial}{\partial t}
\frac{\mathcal{F}_{\pi^0\gamma^*\gamma^*}(-t\mu^2,-t\mu^2)}{\mathcal{F}_{\pi^0\gamma^*\gamma^*}(0,0)}\,.
\label{eq:chiDor}
\end{equation}
Using this approach we get very simple formulae for the VMD model
\begin{equation}
\chi_\text{VMD}^\text{(r)}(\mu)
\simeq\frac{11}4-\frac32\log\frac{ M_{V_1}^2}{\mu^2}
\label{eq:chiVMD}
\end{equation}
as well as for the LMD case (cf.~\cite{Knecht:1999gb})
\begin{equation}
\chi_\text{LMD}^\text{(r)}(\mu)
\simeq\frac{11}4-\frac32\log\frac{ M_{V_1}^2}{\mu^2}
-\frac{12\pi^2F^2}{N_c M_{V_1}^2}\,.
\label{eq:chiLMD}
\end{equation}
For the THS form factor, where the additional mass scale $ M_{V_2}$ and the parameter $\kappa$ come into play, the calculation according to (\ref{eq:chiDor}) yields
\begin{equation}
\begin{split}
&\chi_\text{THS}^\text{(r)}(\mu)
\simeq\frac54-\frac32\log\frac{ M_{V_1}^2}{\mu^2}
+\frac32\frac{ M_{V_1}^2 M_{V_2}^2}{( M_{V_2}^2- M_{V_1}^2)^2}\\
&\times\bigg\{\frac{ M_{V_2}^2}{ M_{V_1}^2}+\frac{ M_{V_1}^2}{ M_{V_2}^2}-\frac{14}{N_c}(2\pi F)^2\bigg(\frac1{ M_{V_1}^2}+\frac1{ M_{V_2}^2}\bigg)\\
&-\bigg[\frac{ M_{V_1}^2}{ M_{V_2}^2}+\frac{2 M_{V_1}^2}{ M_{V_2}^2- M_{V_1}^2}-\frac7{N_c}\frac{(4\pi F)^2}{( M_{V_2}^2- M_{V_1}^2)}\bigg]\log\frac{ M_{V_2}^2}{ M_{V_1}^2}\\
&-\frac{\kappa M_{V_1}^2 M_{V_2}^2}{N_c(4\pi F)^4}\bigg[\frac{ M_{V_2}^2+ M_{V_1}^2}{2( M_{V_2}^2- M_{V_1}^2)}\log\bigg(\frac{ M_{V_2}^2}{ M_{V_1}^2}\bigg)-1\bigg]
\bigg\}\,.
\end{split}
\label{eq:chiTHS}
\end{equation}

The numerical values for the exact as well as for the approximate results are listed in Tab.~\ref{tab:chi}.
\begin{table}[ht]
\begin{center}
{\scriptsize
\begin{tabular}{c | c c c }
\toprule
$\chi^\text{(r)}(\mu=770\,\text{MeV})$ & VMD & LMD & THS\\
\midrule
exact result & 2.87 & 2.29 & $2.2\pm0.2$\\
approx.\ from (\ref{eq:chiDor}) & 2.72 & $2.2\pm0.9$ & $2.0\pm0.2$\\
\bottomrule
\end{tabular}}
\end{center}
\caption{
Values of the effective parameter $\chi^\text{(r)}(770\,\text{MeV})$ for various pion transition form factors.
In general we provide uncertainty estimates only for THS.
The uncertainty for LMD stems from \cite{Knecht:1999gb}.
}
\label{tab:chi}
\end{table}
In the first row the values gained from the matching of (\ref{eq:PLO}) to (\ref{eq:PLOconst}) are listed.
The second row contains the results of the approximative formulae (\ref{eq:chiVMD}), (\ref{eq:chiLMD}) and (\ref{eq:chiTHS}).
In the LMD case, the value according to (\ref{eq:chiLMD}) would be 2.16, but here the listed result has been taken from~\cite{Knecht:1999gb} where also a 40\,\% uncertainty has been estimated.
Needless to say, the used numerical inputs were the same as in the whole text except for one thing.
In the case of the pion decays, we use for the ground-state vector-meson multiplet mass $M_{V_1}$ the average of the $\rho$ and $\omega$ meson masses, i.e.\ $M_{V_1}=(779\pm4)$\,MeV.

We can see that the approximative formula (\ref{eq:chiDor}) is indeed reasonable.
The numerical values are close to the case of the exact calculation, where electron and pion masses are not neglected.
To compare briefly the values for the various approaches, we see in Tab.~\ref{tab:chi} that the THS mean value is close to LMD but incompatible with the VMD value.
Of course, a meaningful comparison would require an uncertainty estimate for the VMD case, which is not available.
Considering the large uncertainty of LMD according to \cite{Knecht:1999gb}, THS and even VMD lie safely within the LMD band.
In turn the central value of LMD lies within the THS band.

Finally, we can utilize the above formulae and provide branching ratios to be compared to experiment.
The presently most precise measurement was performed by the KTeV collaboration at Fermilab with the result~\cite{Abouzaid:2006kk}
\begin{equation}
\begin{split}
&B^\text{KTeV}(\pi^0\to e^+e^-(\gamma)\,,\;x>0.95)\\
&=(6.44\pm0.25\pm0.22)\times10^{-8}\,,
\label{eq:BKTeV}
\end{split}
\end{equation}
where $x$ is the normalized lepton-pair invariant mass, i.e.\ $x=(q_1+q_2)^2/M_\pi^2$.
This condition can be translated to the photon-energy cut $E_\gamma<M(1-x_\text{cut})/2$ in the rest frame of the decaying pion, 
with $x_\text{cut} = 0.95$.
Soon after, the disagreement of (\ref{eq:BKTeV}) with a theoretical calculation was found~\cite{Dorokhov:2007bd}.

To predict the quantity (\ref{eq:BKTeV}) we proceed as follows.
Taking into account NLO QED corrections \cite{Vasko:2011pi,Husek:2014tna} we can calculate the branching ratio of the inclusive process $\pi^0\to e^+e^-(\gamma)$, i.e.\ of a process where we allow bremsstrahlung photons to appear in the final state.
Denoting these QED corrections by $\delta$ this can be written as
\begin{equation}
\begin{split}
&B(\pi^0\to e^+e^-(\gamma)\,,\;x>x_\text{cut})\\
&=\frac{\Gamma^\text{LO}(\pi^0\to e^+e^-)[1+\delta(x_\text{cut})]}{\Gamma^\text{LO}(\pi^0\to \gamma\gamma)}\,
\times B(\pi^0\to \gamma\gamma)\,.
\end{split}
\end{equation}
In addition, the width of the pion decay into two photons has the form 
\begin{equation}
\Gamma^\text{LO}(\pi^0\to \gamma\gamma)
=\frac12\frac1{16\pi M_\pi}\frac{e^4M_\pi^4}{2}
|\mathcal{F}_{\pi^0\gamma^*\gamma^*}(0,0)|^2
\end{equation}
and $B(\pi^0\to \gamma\gamma)=(98.823\pm0.034)\,\%$~\cite{PDG}.
The calculation of the two-loop virtual radiative corrections together with the bremsstrahlung correction in the soft-photon approximation gives the result $\delta(0.95)=(-5.8\pm0.2)\,\%$.
It has been shown in~\cite{Husek:2014tna} that the soft-photon approximation is a valid approach for the value $x_\text{cut}$=0.95 used by the KTeV experiment.
Note that this value of the correction $\delta$ was obtained in~\cite{Vasko:2011pi} in a model-independent way using $\chi^\text{(r)}(770\,\text{MeV})=2.2\pm0.9$.
However, this result for $\delta$ depends negligibly on the range of considered values of $\chi^\text{(r)}$ shown in Tab.~\ref{tab:chi}.
Hence we consider the result for $\delta$ to be valid for our case.

With this NLO QED input we find the values shown in Tab.~\ref{tab:KTeV}.
\begin{table}[!ht]
\begin{center}
{\scriptsize
\begin{tabular}{c | c c c c }
\toprule
& VMD & LMD & THS & KTeV\\
\midrule
$B(\pi^0\to e^+e^-(\gamma)\,,\;x>0.95)$ & 5.96 & 5.8(3) & 5.76(7) & 6.44(33)\\
\bottomrule
\end{tabular}}
\end{center}
\caption{
Branching ratio of the inclusive process $\pi^0\to e^+e^-(\gamma)$ at NLO of QED for various models for the pion transition form factor.
The listed values are to be multiplied by a factor $10^{-8}$.
The KTeV value is based on the result stated in~\cite{Abouzaid:2006kk}.
}
\label{tab:KTeV}
\end{table}
Here we see that the central value in the THS case is, of course, compatible with the LMD central value, since the same holds for the values $\chi^\text{(r)}(\mu)$; see Tab.~\ref{tab:chi}.
The uncertainty of the LMD approach according to \cite{Knecht:1999gb} makes it compatible with the KTeV result on the level of 1\,$\sigma$.
Considering VMD, where we do not have an estimate of the theoretical uncertainty, we are then 1.5\,$\sigma$ away from the experimental value.
Finally, the THS approach is 2\,$\sigma$ off.
As compared to LMD this is related to the fact that the estimate for our theoretical uncertainty is on the level of only 1\,\% concerning this particular branching ratio.
We recall that this estimate is merely based on two sources: first, the fit for $\kappa$, which might be improved in the future by data on the doubly virtual pion transition form factor; second, a large variation in $M_{V_2}$.
By this variation we estimate the uncertainty imposed by a truncation of the infinite tower of resonances.
Strictly speaking our model intrinsic uncertainty estimate does not account for the two initial assumptions: the chiral and the large-$N_c$ limit.
We expect that to some extent we have accounted for the uncertainties of the large-$N_c$ approximation by focusing on kinematical situations where the (large-$N_c$ suppressed) widths of the resonances do not matter so much {\em and} by varying the mass of the second vector-meson multiplet in a relatively large range.
The uncertainties caused by the chiral limit should be on the order of $M_\pi^2/\Lambda_\chi^2$ where $\Lambda_\chi$ denotes the scale of chiral symmetry breaking.
Within chiral perturbation theory one uses the scale where new degrees of freedom come into play or where the loops become as important as the tree-level contributions.
Roughly this implies $\Lambda_\chi \lesssim M_\rho, 4\pi F$, i.e.\ an uncertainty of about 3\,\%.
Adding errors in squares this is clearly negligible as compared to the 10\,\%\ uncertainty documented in Tab.~\ref{tab:chi} for THS.
Based on these considerations we regard our uncertainty estimate in Tab.~\ref{tab:KTeV} as reasonable --- assuming that our determination of $\kappa$ from the $\omega$-$\pi$ transition is correct.

Consequently, if the (central value of the) KTeV result will be confirmed by future experiments and measured with higher precision such that the discrepancy would reach several $\sigma$'s, then the following two scenarios are conceivable: 
Either there are some aspects of the THS approach which are not well-suited for the rare decay $\pi^0\to e^+e^-$, or it should be seriously considered that physics beyond the Standard Model influences the rare pion decay to a significant extent.
However, under the present circumstances we consider the current discrepancy to be inconclusive.
We will come back to this discrepancy in the next section.


\section{Singly virtual pion transition form factor and the Dalitz decay \texorpdfstring{$\pi^0\to e^+e^-\gamma$}{pi0 --> e+e-gamma}}
\label{sec:pieegam}

With particular models at hand, we can also explore the properties of the singly virtual pion transition form factor $\mathcal{F}_{\pi^0\gamma^*\gamma^*}(0,q^2)$.
A slope $a_\pi$ and a curvature $b_\pi$ of the form factor are defined in terms of the Taylor expansion in the invariant mass of the vector current \cite{Hoferichter:2014vra}
\begin{equation}
\frac{\mathcal{F}_{\pi^0\gamma^*\gamma^*}(0,q^2)}{\mathcal{F}_{\pi^0\gamma^*\gamma^*}(0,0)}
\simeq1+a_\pi\frac{q^2}{M_\pi^2}+b_\pi\bigg(\frac{q^2}{M_\pi^2}\bigg)^2\,.
\end{equation}
By means of the first and second derivative it is straightforward to find the following expressions for the simplest form factors
\begin{equation}
a_\pi^\text{VMD}=\frac{M_\pi^2}{ M_{V_1}^2}\;,\quad
b_\pi^\text{VMD}=\frac{M_\pi^2}{ M_{V_1}^2}\,a_\pi^\text{VMD}\,,
\end{equation}
\begin{equation}
a_\pi^\text{LMD}
=\frac{M_\pi^2}{ M_{V_1}^2}\bigg[1-\frac{4\pi^2 F^2}{N_c  M_{V_1}^2}\bigg]\;,\quad
b_\pi^\text{LMD}
=\frac{M_\pi^2}{ M_{V_1}^2}\,a_\pi^\text{LMD}\,
\end{equation}
as well as for the THS form factor with its two vector-multiplet mass scales
\begin{equation}
\begin{split}
a_\pi^\text{THS}
&=\frac{M_\pi^2}{ M_{V_1}^2}\bigg[1+\frac{ M_{V_1}^2}{ M_{V_2}^2}-\frac{24\pi^2 F^2}{N_c  M_{V_2}^2}\bigg]\;,\quad\\
b_\pi^\text{THS}
&=\bigg[\frac{M_\pi^2}{ M_{V_1}^2}+\frac{M_\pi^2}{ M_{V_2}^2}\bigg]\,a_\pi^\text{THS}
-\frac{M_\pi^4}{ M_{V_1}^2 M_{V_2}^2}\,.
\end{split}
\label{eq:abTHS}
\end{equation}
Numerical results are shown in Tab.~\ref{tab:slope}.
\begin{table}[!ht]
\begin{center}
{\scriptsize
\begin{tabular}{c | c c c c c }
\toprule
& VMD & LMD & THS & dispers.~\cite{Hoferichter:2014vra} & exp.~\cite{PDG}\\
\midrule
slope $a_\pi$ & 30.0 & 24.5 & $29.2(4)$ & 30.7(6) & $32(4)$\\
curvature $b_\pi$ & 0.90 & 0.74 & $0.87(2)$ & 1.10(2) & --\\
\bottomrule
\end{tabular}}
\end{center}
\caption{
Slope and curvature of the singly virtual pion transition form factor evaluated for various approaches.
The listed values are to be multiplied by a factor of $10^{-3}$.
}
\label{tab:slope}
\end{table}
We see that THS is compatible with the experimental value and that it is numerically consistent with VMD.
Similar to some previous cases, the value predicted by LMD is unsatisfying.
Nevertheless, none of the models (VMD, LMD, THS) fully agrees with the results of the dispersive calculation \cite{Hoferichter:2014vra}.
The latter is capable of producing very reliable results at low energies.
We note in passing that the apparent agreement of VMD with the dispersive result for the slope is somewhat accidental.
Actually the discrepancy grows, if the $\phi$ meson is included in the analysis \cite{Hoferichter:2014vra}.

The disagreement between THS and the dispersive results brings us back to the discussion of the uncertainties inherent to THS and to our attempt to quantify these uncertainties in a reasonable way.
Of course, for a given model one can only estimate the model intrinsic uncertainties, which in turn inherit a model dependence.
Our numerical values for the uncertainty estimates of the THS results emerge dominantly from the uncertainty in the determination of the parameter $\kappa$ and from a variation in $M_{V_2}$.
In contrast to most of the results previously presented, the singly virtual pion transition form factor does not depend on the parameter $\kappa$.
This can be most easily seen from (\ref{eq:abTHS}), but also from (\ref{eq:defpiTFF}), (\ref{eq:FFfinal2}).
In addition, the dependence on $M_{V_2}$ is minor for the low-energy quantities determined in (\ref{eq:abTHS}).
In fact, in (\ref{eq:abTHS}) the mass $M_{V_2}$ essentially shows up in the numerically small combination $\Delta M^2/M_{V_2}^2$ with $\Delta M^2 \equiv M_{V_1}^2-24 \pi^2 F^2/N_c$.
The smallness of $\Delta M^2$ might be seen as an incarnation of the KSFR relations; see, e.g., \cite{Leupold:2003zb} and references therein.
In turn, this implies that our model intrinsic uncertainty determination might underestimate the real uncertainty if the dependence of our results on $M_{V_2}$ and/or $\kappa$ is accidentally small.
While this is the case for the singly virtual pion transition form factor, it is certainly not true for the doubly virtual one; see also Fig.~\ref{fig:FF} with the broad uncertainty band in the first panel and the corresponding nearly invisible spread in the second panel.
Obviously, one has to take our uncertainty estimates with a grain of salt.

\begin{figure}[t]
\centering
\begin{subfigure}[t]{0.3\columnwidth}
\centering
\setlength{\unitlength}{0.4pt}
  \begin{picture}(100,250) (220,-250)
    \SetScale{0.4}
    \SetWidth{1.0}
    \SetColor{Black}
    \Line[dash,dashsize=10](180,-123)(230,-123)
    \Photon(234,-123)(306,-177){4}{5.5}
    \Line[arrow,arrowpos=0.5,arrowlength=5,arrowwidth=2,arrowinset=0.2](342,-123)(288,-87)
    \Line[arrow,arrowpos=0.5,arrowlength=5,arrowwidth=2,arrowinset=0.2](288,-87)(342,-51)
    \Photon(234,-123)(288,-87){4}{4.5}
    \GOval(245,-123)(14,14)(0){0.882}
  \end{picture}
\caption{}
\label{fig:D}
\end{subfigure}
\begin{subfigure}[t]{0.28\columnwidth}
\centering
\setlength{\unitlength}{0.4pt}
\begin{picture}(170,150) (180,-190)
    \SetScale{0.4}
    \SetWidth{1.0}
    \SetColor{Black}
    \Line[dash,dashsize=10](180,-53)(234,-53)
    \Photon(234,-53)(306,1){4}{5.5}
    \Line[arrow,arrowpos=0.5,arrowlength=5,arrowwidth=2,arrowinset=0.2](360,-143)(306,-107)
    \Line[arrow,arrowpos=0.5,arrowlength=5,arrowwidth=2,arrowinset=0.2](306,-107)(306,1)
    \Line[arrow,arrowpos=0.5,arrowlength=5,arrowwidth=2,arrowinset=0.2](342,25)(378,49)
    \Photon(306,-107)(234,-53){4}{5.5}
    \Photon(342,25)(378,1){4}{3.5}
    \Line[arrow,arrowpos=0.5,arrowlength=5,arrowwidth=2,arrowinset=0.2](306,1)(342,25)
    \Text(210,-170)[lb]{\large{\Black{+ cross}}}
    \GOval(250,-53)(18,18)(0){0.882}
\end{picture}
\caption{}
\label{fig:1gIRa}
\end{subfigure}
\begin{subfigure}[t]{0.33\columnwidth}
\centering
\setlength{\unitlength}{0.4pt}
\begin{picture}(180,150) (160,-210)
    \SetScale{0.4}
    \SetWidth{1.0}
    \SetColor{Black}
    \Line[dash,dashsize=10](180,-88)(234,-88)
    \Photon(234,-88)(306,-34){4}{5.5}
    \Line[arrow,arrowpos=0.5,arrowlength=5,arrowwidth=2,arrowinset=0.2](360,-178)(306,-142)
    \Line[arrow,arrowpos=0.5,arrowlength=5,arrowwidth=2,arrowinset=0.2](306,-88)(306,-34)
    \Line[arrow,arrowpos=0.5,arrowlength=5,arrowwidth=2,arrowinset=0.2](306,-34)(360,2)
    \Photon(306,-142)(234,-88){4}{5.5}
    \Line[arrow,arrowpos=0.5,arrowlength=5,arrowwidth=2,arrowinset=0.2](306,-142)(306,-88)
    \Photon(306,-88)(366,-88){4}{4}
    \GOval(250,-88)(18,18)(0){0.882}
\end{picture}
\caption{}
\label{fig:1gIRb}
\end{subfigure}
\caption{
Leading order diagram of the Dalitz decay $\pi^0\to e^+e^-\gamma$ in the QED expansion (a) and the one-photon irreducible contributions to the NLO virtual radiative correction (b) and (c).
Note that ``cross'' in (b) accounts for a diagram with a photon emitted from the outgoing positron line.
Diagrams (b) and (c) also serve as a bremsstrahlung contribution to the $\pi^0\to e^+e^-$ decay.
}
\label{fig:D1gIR}
\end{figure}
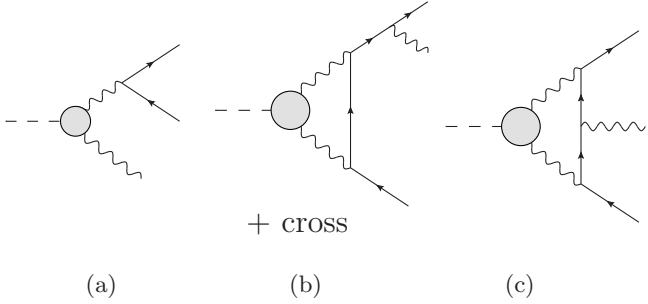

As a closure of the phenomenological part of this work, we inspect the theoretical predictions for the decay width of the Dalitz decay $\pi^0\to e^+e^-\gamma$.
The leading-order QED contribution is depicted in Fig.~\ref{fig:D}.
Our aim is also to address the NLO radiative corrections.
For this purpose we use the approach documented in \cite{Husek:2015sma}, which has recently reviewed and extended the classical work of~\cite{Mikaelian:1972yg}.
Hence, together with the bremsstrahlung beyond the soft-photon approximation we use in the following calculations the virtual radiative corrections including also the one-photon irreducible (1$\gamma$IR) contribution.
The diagrams connected with the latter part are schematically shown in Fig.~\ref{fig:1gIRa} and ~\ref{fig:1gIRb}.
The mentioned correction then emerges when one considers the interference term with the LO diagram in Fig.~\ref{fig:D}.

The decay width of the Dalitz decay at NLO can be written in the following form
\begin{equation}
\begin{split}
&\Gamma(\pi^0\to e^+e^-\gamma(\gamma)\,,\;x>x_\text{cut})\\
&=[1+\delta_D(x>x_\text{cut})]\,\Gamma^\text{LO}(\pi^0\to e^+e^-\gamma\,,\;x>x_\text{cut})\,,
\end{split}
\label{eq:dalitzNLOSL}
\end{equation}
where we have denoted the sum of all the NLO radiative corrections to the integrated decay width as $\delta_D$ and where
\begin{equation}
\begin{split}
&\Gamma^\text{LO}(\pi^0\to e^+e^-\gamma\,,\;x>x_\text{cut})\\
&=\Big(\frac{\alpha}{\pi}\Big)\,\Gamma^\text{LO}(\pi^0\to\gamma\gamma)
\int_{x_\text{cut}}^1\hskip-2pt\text{d}x\,
\bigg|\frac{\mathcal{F}_{\pi^0\gamma^*\gamma^*}(0,xM_\pi^2)}{\mathcal{F}_{\pi^0\gamma^*\gamma^*}(0,0)}\bigg|^2\\
&\times\frac{8\beta_e(x)}{3}\frac{(1-x)^3}{4x}\bigg(1+\frac{2m_e^2}{xM_\pi^2}\bigg)\,.
\label{eq:dalitzXLO}
\end{split}
\end{equation}
In the above formula we have introduced $\beta_e(x)=\sqrt{1-{4m_e^2}/{(xM_\pi^2)}}$.
Note that contrary to the last section $x_\text{cut}$ cannot be interpreted as a cut on a single photon energy, since the additional bremsstrahlung photon appears.
There are also no restrictions on the energy of the bremsstrahlung photon in this decay width.

The numerical results for the considered approaches are compared in Tab.~\ref{tab:Dalitz}.
\begin{table}[!ht]
\begin{center}
{\tiny
\begin{tabular}{c | c c c c }
\toprule
& WZW & VMD & LMD & THS\\
\midrule
NLO correction $\delta_\text{D}$ & 8.338 & 8.299 & 8.307 & 8.3008(3)\\
$\frac{\Gamma(\pi^0\to e^+e^-\gamma(\gamma))}{\Gamma^\text{LO}(\pi^0\to \gamma\gamma)}$ & 11.9503 & 11.9735 & 11.9692 & 11.9729(4)\\
$\frac{\Gamma(\pi^0\to e^+e^-\gamma(\gamma)\,,\;x>0.2319)}{\Gamma(\pi^0\to e^+e^-\gamma(\gamma))}$ & 31.03 & 31.63 & 31.52 & 31.61(1)\\
\bottomrule
\end{tabular}}
\end{center}
\caption{
Theoretical prediction for the Dalitz decay branching ratios at NLO of QED with and without a cut as evaluated for various approaches.
``WZW'' describes the case where the pion transition form factor is replaced by its low-energy limit.
The listed values are to be multiplied by a factor of $10^{-3}$.
}
\label{tab:Dalitz}
\end{table}
In the first row the NLO correction is presented.
Naively one might get from the definitions (\ref{eq:dalitzNLOSL}) and (\ref{eq:dalitzXLO}) the impression that the pion transition form factor does not enter $\delta_D$.
This is, however, not the case.
The NLO correction involves, of course, the four-body phase-space integrations and the pion transition form factor cannot be scaled out.
This is due to the integration over the $x$ variable, which is tacitly performed also on the left-hand side of (\ref{eq:dalitzNLOSL}).
On the other hand, taking into account the correction to the differential decay width, we would get an expression independent of the form factor.
This statement then holds only when leaving out the 1$\gamma$IR part, where the form-factor-dependent loop integration is performed.
Nonetheless, concerning model uncertainties, it is encouraging that numerically the NLO correction is to a very large extent independent of the used pion transition form factor.
Using the respective value of the NLO correction, the second row in Tab.~\ref{tab:Dalitz} shows the corresponding branching ratio if no cut on additional photons is applied, i.e.\ the branching ratio of the Dalitz decay for the whole kinematic region $x\in[4m_e^2/M_\pi^2,1]$.
The results of this row should be compared to the experimental value $\frac{B(\pi^0\to e^+e^-\gamma)}{B(\pi^0\to \gamma\gamma)}=(11.88\pm0.35)\times10^{-3}$, which has been calculated using inputs from~\cite{PDG}.
The agreement with the values of the second row in Tab.~\ref{tab:Dalitz} is very satisfying.

In the previous section we have stated the value (\ref{eq:BKTeV}) as the outcome of the KTeV analysis.
However, what KTeV has determined at first place is%
\footnote{Note that we use here the more precise value $x_\text{cut}=0.2319$ from~\cite{Rune:2006} in contrast to 0.232 as stated in~\cite{Abouzaid:2006kk}.
This value for $x_\text{cut}$ corresponds to (65\,MeV/$M_\pi$)$^2$.}
\begin{equation}
\begin{split}
&\frac{\Gamma(\pi^0\to e^+e^-(\gamma)\,,\;x>0.95)}{\Gamma(\pi^0\to e^+e^-\gamma(\gamma)\,,\;x>0.2319)} \Bigg\vert_\text{KTeV} \\
&=(1.685\pm0.064\pm0.027)\times10^{-4}   \,.
\end{split}
\label{eq:KTeV232}
\end{equation}
We note in passing that for the quantity displayed in the last row of Tab.~\ref{tab:Dalitz} the corresponding value used by KTeV to obtain finally the result (\ref{eq:BKTeV}) is 3.19\,\%~\cite{Abouzaid:2006kk}.
This value is close but not equal to any of the results in the last row of Tab.~\ref{tab:Dalitz}.

Having the radiative corrections determined from theory suggests to calculate directly the ratio (\ref{eq:KTeV232}) instead of the derived quantity (\ref{eq:BKTeV}) and to compare the calculation to the experimental result.
However, the result (\ref{eq:KTeV232}) is not a direct measurement either, but is, of course, based on an interplay of measurements, Monte Carlo simulations, acceptance corrections, etc.\ \cite{Abouzaid:2006kk,Rune:2006}.
By the time the KTeV measurements were analyzed not all QED radiative corrections were available.
To evaluate the decay width of the Dalitz decay the radiative corrections from~\cite{Mikaelian:1972yg} have been used.
In particular the 1$\gamma$IR contributions have not been included in \cite{Mikaelian:1972yg}.
In addition, in the Monte Carlo generator of KTeV the radiative corrections for the $e^+ e^-$ decay have been adopted from \cite{Bergstrom:1982wk}; see \cite{Rune:2006}.
However, in \cite{Bergstrom:1982wk} some approximations have been used that have been proven to be misleading after the exact calculation has been performed~\cite{Vasko:2011pi}.
Therefore it would be somewhat misleading to compare a theory result with all radiative corrections included to an experimental result where only part of these corrections have been taken into account.
On the other hand, it might be illuminating to provide a rough estimate of the impact of these differences.
Before presenting such an estimate we would like to repeat the take-home message from the first section:
All NLO QED radiative corrections are now available \cite{Vasko:2011pi,Husek:2014tna,Husek:2015sma} and can be taken into account in future analyses of data on the pion decays $\pi^0 \to e^+ e^-$ and $\pi^0 \to e^+ e^- \gamma$.

In spite of the fact that not all radiative corrections have been taken into account in the analysis that led to the result (\ref{eq:KTeV232}) we do not feel legitimated to modify this experimental result.
Instead we will calculate the ratio of (\ref{eq:KTeV232}) within THS%
\footnote{According to Tab.~\ref{tab:Dalitz} the results from the other approaches are comparable.},
but assign an uncertainty to it, which is related to the neglect of the 1$\gamma$IR contributions.%
\footnote{The modification induced by using \cite{Bergstrom:1982wk} in the Monte Carlo generator is even harder to assess.
Therefore we concentrate solely on the 1$\gamma$IR contributions.}
It should be stressed, however, that this is not an entirely valid approach either.
After all, as soon as one can calculate the NLO QED radiative corrections, they do not constitute an uncertainty of a result but rather a well-defined shift of the result.
Nonetheless, we proceed in the described way to obtain a rough estimate of the uncertainty caused by neglecting part of the radiative corrections.

The quantity of our interest can be generally expressed and expanded as
\begin{equation}
\frac{A\pm\sigma(A)}{B\pm\sigma(B)}
\simeq\frac AB\left[1\pm\frac{\sigma(A)}{A}\mp\frac{\sigma(B)}{B}\right]\,.
\end{equation}
In our case we have $A=\Gamma(\pi^0\to e^+e^-(\gamma)\,,\;x>0.95)$ and $B=\Gamma(\pi^0\to e^+e^-\gamma(\gamma)\,,\;x>0.2319)$, which we calculate up to NLO using the complete set of corrections.
The uncertainties are then given by the 1$\gamma$IR contributions to the NLO virtual radiative corrections, i.e.\ by the interference of the diagrams shown in Fig.~\ref{fig:D1gIR}.
Hence
\begin{equation}
\begin{split}
\frac{\sigma(A)}{A}
&=\frac{\big|\Gamma_{1\gamma\text{IR}}^\text{NLO}(\pi^0\to e^+e^-\gamma\,,\;x>0.95)\big|}{\Gamma(\pi^0\to e^+e^-(\gamma)\,,\;x>0.95)}\\
&=\frac{\big|\delta_\text{D}^{1\gamma\text{IR}}(x>0.95)\big|}{[1+\delta(0.95)]}\frac{\Gamma^\text{LO}(\pi^0\to e^+e^-\gamma\,,\;x>0.95)}{\Gamma^\text{LO}(\pi^0\to e^+e^-)}   \,,
\label{eq:sA}
\end{split}
\end{equation}
which numerically yields $\sigma(A)/A \simeq0.4\,\%$.
For the second term we get
\begin{equation}
\begin{split}
\frac{\sigma(B)}{B}
&=\frac{\big|\Gamma_{1\gamma\text{IR}}^\text{NLO}(\pi^0\to e^+e^-\gamma\,,\;x>0.2319)\big|}{\Gamma(\pi^0\to e^+e^-\gamma(\gamma)\,,\;x>0.2319)}\\
&=\frac{\big|\delta_\text{D}^{1\gamma\text{IR}}(x>0.2319)\big|}{[1+\delta_D(x>0.2319)]}\,,
\label{eq:sB}
\end{split}
\end{equation}
for which we find $\sigma(B)/B \simeq0.5\,\%$.
We can thus conclude that the total uncertainty would be at the level of 1\,\%.
For completeness we list the values that we have used to get the above stated numbers.
The input for (\ref{eq:sA}) is $\delta_\text{D}^{1\gamma\text{IR}}(x>0.95)=-8.93(1)\,\%$ for the 1$\gamma$IR contribution to the NLO virtual radiative correction for the decay width of the Dalitz decay.
For the ratio of the LO decay widths of the Dalitz decay with $x>0.95$ and the $\pi^0\to e^+e^-$ process we have $\frac{\Gamma^\text{LO}(\pi^0\to e^+e^-\gamma\,,\;x>0.95)}{\Gamma^\text{LO}(\pi^0\to e^+e^-)}=4.31(5)\,\%$.
In (\ref{eq:sB}) we have utilized for the overall NLO correction to the Dalitz decay 
$\delta_\text{D}(x>0.2319)=-3.189(2)\,\%$ and finally $\delta_\text{D}^{1\gamma\text{IR}}(x>0.2319)=-0.5038(3)\,\%$.
In general, we can see that the radiative corrections are getting more important with higher values of $x_\text{cut}$.
Taking values from Tab.~\ref{tab:KTeV} and Tab.~\ref{tab:Dalitz} and the previously estimated relative uncertainty of 1\,\% (we state this separately) we find
\begin{equation}
\begin{split}
&\frac{\Gamma(\pi^0\to e^+e^-(\gamma)\,,\;x>0.95)}{\Gamma(\pi^0\to e^+e^-\gamma(\gamma)\,,\;x>0.2319)} \Bigg\vert_\text{theo+uncert.} \\
&=(1.54\pm0.02\pm0.02)\times10^{-4}\,.
\end{split}
\label{eq:finallabel}
\end{equation}
This result is to be compared with the KTeV value (\ref{eq:KTeV232}), yielding a discrepancy of approximately 1.5\,$\sigma$.
We stress one more time that the result (\ref{eq:finallabel}) should be read with caution.


\section{Outlook}
\label{sec:outlook} 

Several directions are conceivable how the THS scheme might be extended or combined with other approaches.
Most straightforward would be to extend THS to other correlators; see also~\cite{Knecht:2001xc}.
Eventually one might get cross correlations and in that way more information, e.g., by fitting to additional sets of data.

In the present work we have merely used the $PVV$ correlator as an intermediate step to obtain the $\pi VV$ correlator.
Of course, the $PVV$ correlator is also interesting in its own right.
For instance, as it has been worked out in \cite{Knecht:2001xc}, the low-energy expansion of this correlator determines some of the low-energy constants of the chiral Lagrangian at order $p^6$ in the anomalous sector \cite{Ebertshauser:2001nj}, i.e.\ the corrections to the Wess--Zumino--Witten Lagrangian.
Having determined some of the parameters of the $PVV$ correlator within THS by demanding high-energy constraints one could provide estimates and/or cross-correlations for these low-energy constants following the procedure outlined in \cite{Knecht:2001xc}.

As already spelled out in Section~\ref{sec:intro} we refrained from using high-energy constraints for quantities like the $\pi \omega V$ correlator since the vector meson might probe already too many details of the intermediate-energy region (above 1 GeV) where we have approximated the tower of infinitely many states by one effective state per channel.
Extending the scheme from THS to three hadrons per channel, albeit introducing more parameters in the first place, might allow to use quark-scaling relations along the lines of \cite{Lepage:1979zb,Lepage:1980fj,Brodsky:1981rp} for $\pi\omega V$ and similar correlators which in turn could help to keep the number of free parameters manageable.
In the same spirit one can use subleading orders in the high-energy expansion in terms of QCD condensates \cite{Novikov:1983jt}.

In the present work we have used the chiral limit to construct the $PVV$ correlator.
In line with this approximation we have focused on the $\pi VV$ and $\pi \omega V$ correlator where no strange quarks are involved.
For a reasonable extension from the pion to the whole pseudoscalar multiplet (a nonet in the large-$N_c$ limit) one has to go beyond the chiral limit.
The prospect of such an endeavor would be the possibility to tackle the transition form factors of the $\eta$ and $\eta'$ mesons and their rare decays into electron and positron or muon and anti-muon.
The simplest way to go beyond the chiral limit would be to utilize a pertinent Lagrangian.
In extensions of chiral perturbation theory~\cite{Weinberg:1978kz,Gasser:1983yg,Gasser:1984gg} one frequently uses a Lagrangian where the vector mesons are represented by antisymmetric tensor fields~\cite{Gasser:1983yg,Ecker:1988te,Kampf:2011ty,Terschluesen:2010ik,Terschlusen:2012xw,Terschlusen:2013iqa,Roig:2013baa,Roig:2014uja}.
As discussed in~\cite{Knecht:2001xc} a given resonance Lagrangian of formal chiral order $p^6$ might not be capable of satisfying all desired OPE constraints.
Indeed, as shown in~\cite{Kampf:2011ty} this problem appears at least if only one vector-meson multiplet is used for the $PVV$ correlator.
Concerning the vector instead of the antisymmetric tensor representation the same negative result has been reported in~\cite{Knecht:2001xc}.
It might appear that the first-order approach suggested in~\cite{Kampf:2006yf} offers more flexibility here.
Alternatively or in addition one might consider the inclusion of resonance terms of chiral order $p^8$.
The following consideration might demonstrate that one can expect to create additional terms that would not spoil the OPE right away: If one normalizes the $PVV$ correlator to the corresponding chiral-anomaly expression, a $p^6$ resonance Lagrangian
would produce ``corrections'' that schematically are given by
\begin{equation}
{\mathcal L}_{p^6,R}\quad\to\quad\frac{Q^2}{\left(Q^2+M_R^2\right)^n}\quad\text{with}\quad n \le 3 \,.
\label{eq:schemat}
\end{equation}
We have replaced all resonance masses by $M_R$ and all momenta by $Q$.
This is sufficient to discuss the high- and low-energy limits.
The factor of $Q^2$ in the numerator of \eqref{eq:schemat} emerges from the fact that the resonance Lagrangian is assumed to be of chiral order $p^6$ while the chiral anomaly is of order $p^4$.
The limitation $n \le 3$ comes from the fact that a three-point correlator is considered.
At high energies a single term with $n=1$ definitely spoils the OPE.
Coupling constants between various contributions must be adjusted such that cancellations appear.
Of course, this is the reason why the OPE is capable to tell something about the resonance parameters.
A resonance Lagrangian of chiral order $p^8$ produces terms with the schematic structure
\begin{equation}
{\mathcal L}_{p^8,R}\quad\to\quad\frac{Q^4}{\left(Q^2+M_R^2\right)^n}\quad\text{with}\quad n \le 3 \,.
\label{eq:schemat2}
\end{equation}
Here the terms with $n=1,2$ spoil the OPE and must cancel.
The terms with $n=3$, however, produce structures that could be compatible with the OPE but are structurally different from the ones in \eqref{eq:schemat}.
Thus the inclusion of resonance terms of chiral order $p^8$ might help to obtain an agreement between a Lagrangian approach and the OPE.
Needless to say that the systematic construction of a $p^8$ Lagrangian would be extremely tedious; see~\cite{Kampf:2011ty} and references therein.
In practice, according to \cite{Mateu:2007tr}, leading-order OPE constraints%
\footnote{Note that ``chiral order'' refers to the low-energy expansion, while ``leading-order OPE'' refers to the high-energy expansion.}
together with a singly virtual pion transition form factor that vanishes at large momenta can be achieved by a resonance 
Lagrangian with two vector-meson multiplets in the antisymmetric tensor representation.
Note, however, that the concrete B-L limit (\ref{eq:BL}) has not been used in \cite{Mateu:2007tr}.
Nonetheless, the results of \cite{Mateu:2007tr} suggest that THS can be reproduced from a resonance Lagrangian of 
order $p^6$ in the chiral counting.

While leaving the chiral limit might be systematized by a Lagrangian approach, our second basic assumption, the large-$N_c$ limit is much harder to remedy.
On purpose we stuck to narrow states ($\pi$, $\omega$) and to the space-like or low-energy time-like region where the effects of inelasticities are minor.
Nonetheless this induces uncertainties, which are hard to assess on a quantitative level.
At low energies an excellent way to deal with inelasticities is to use them in ones favor by a dispersive setup~\cite{Schneider:2012ez,Hoferichter:2014vra,Ananthanarayan:2014pta}.
A powerful framework emerges when combined with chiral low-energy constraints, QCD high-energy constraints, and precise data on the scattering amplitudes that come into play via the inelasticities.
In practice the scheme becomes intractable at intermediate energies (roughly above 1 GeV) where more and more many-particle channels open up.
Even if one studies low-energy quantities one might need some information from the intermediate-energy regime if the QCD high-energy limit is reached very late.
This is exactly what we saw for the THS result of the $\pi VV$ correlator.
This suggests to study the impact of the THS results on a dispersive calculation by using THS in the intermediate- and high-energy space-like regime where its results should be most trustworthy.

In the present work we have determined the $\pi VV$ correlator with a focus on the rare pion decay into electron and positron.
In turn this decay might offer some window to observe low-energy traces of physics beyond the Standard Model.
However, the $\pi VV$ correlator is also interesting because it contains information about the intrinsic structure of the pion.
Hence it could be illuminating to figure out what the THS result implies for the pion distribution function~\cite{Lepage:1979zb,Lepage:1980fj,Brodsky:1981rp,Cloet:2013tta}.


\section*{Acknowledgment}
T.H.\ would like to thank for the hospitality of Uppsala University and to all his colleagues there who made a very friendly atmosphere.
We also acknowledge valuable discussions with our colleagues K.\ Kampf and J.\ Novotn\'y from Prague.
We are indebted to M.\ Knecht for bringing \cite{Novikov:1983jt} to our attention.

The present work was supported by the Charles University grants GAUK 700214 and SVV 260219/2015, by the Czech Science Foundation grant GA\v{C}R 15-18080S and by the project ``Study of Strongly Interacting Matter'' (HadronPhysics3, Grant Agreement No.~283286) under the 7th Framework Program of the EU.


\appendix

\section{Explicit form factor formulae}
\label{app:FF}

In Section~\ref{sec:piee} we have defined the invariant amplitude for the process $\pi^0\to e^+e^-$ using the expression $P_{\pi^0\to e^+e^-}^\text{LO}$.
In this appendix we summarize its explicit form for various approaches that lead to a rational function for the pion transition form factor.
We make use of the standard Passarino--Veltman \cite{Passarino:1978jh} scalar one-loop integrals $B_0$ and $C_0$.
The only UV divergent function is then $B_0$.
Its explicit form will be given here as a reference point for the used notation:
\begin{equation}
\begin{split}
&i\pi^2 B_0(0,m^2,m^2)
\equiv(2\pi)^4\mu^{4-d}\int\frac{\text{d}^dl}{(2\pi)^d}\frac{1}{\left[
l^2-m^2+i\epsilon\right]^2}\\
&=i\pi^2\Bigl[\frac 
1\varepsilon-\gamma_E+\log4\pi+\log\left(\frac{\mu^2}{m^2}\right)\Bigr]\,,
\end{split}
\end{equation}
where we have introduced $\varepsilon=2-\tfrac{d}{2}$.
Note that differences of $B_0$ functions are finite.
It will then be manifest from the form in which the results are presented, that the divergent parts cancel yielding a finite amplitude as desired.
It is also convenient to introduce the following combination of the three-point scalar one-loop function $C_0$ and the K\"all\'en triangle function $\lambda$:
\begin{equation}
\begin{split}
&C^\prime_0(m_e^2,m_e^2,m_1^2,m_2^2,m_e^2,m_3^2)\\
&\equiv\frac{1}{m_1^2}\lambda(m_1^2,m_2^2,m_3^2)\,C_0(m_e^2,m_e^2,m_1^2,m_2^2,m_e^2,m_3^2) \,.
\end{split}
\end{equation}
For instance, we present two $C^\prime_0$ functions that are used further in this appendix:
\begin{equation}
\begin{split}
&C^\prime_0(m_e^2,m_e^2,M_\pi^2,0,m_e^2, M_{V_1}^2)\\
&=\frac{1}{M_\pi^2}( M_{V_1}^2-M_\pi^2)^2\,C_0(m_e^2,m_e^2,M_\pi^2,0,m_e^2, M_{V_1}^2)\,,
\end{split}
\end{equation}
\begin{equation}
\begin{split}
&C^\prime_0(m_e^2,m_e^2,M_\pi^2, M_{V_1}^2,m_e^2, M_{V_1}^2)\\
&=-(4 M_{V_1}^2-M_\pi^2)\,C_0(m_e^2,m_e^2,M_\pi^2, M_{V_1}^2,m_e^2, M_{V_1}^2)\,.
\end{split}
\end{equation}
Now we can finally list the results for $P_{\pi^0\to e^+e^-}$ as obtained from the various approaches to the pion transition form factor:
\begin{equation}
\begin{split}
&P_{\pi^0\to e^+e^-}^\text{VMD}
=-\frac{e^4m_e}{16\pi^2}
\left(-\frac{N_c}{12\pi^2F}\right)\\
&\times\bigg\{
\frac{2 M_{V_1}^2}{M_\pi^2}\big[B_0(m_e^2,m_e^2, M_{V_1}^2)-B_0(0,m_e^2,m_e^2)-2\big]\\
&+C^\prime_0(m_e^2,m_e^2,M_\pi^2,0,m_e^2,0)\\
&-2\,C^\prime_0(m_e^2,m_e^2,M_\pi^2,0,m_e^2, M_{V_1}^2)\\
&+C^\prime_0(m_e^2,m_e^2,M_\pi^2, M_{V_1}^2,m_e^2, M_{V_1}^2)
\bigg\}\,,
\end{split}
\end{equation}
\begin{equation}
\begin{split}
&P_{\pi^0\to e^+e^-}^\text{LMD}
=P_{\pi^0\to e^+e^-}^\text{VMD}
-\frac{e^4m_e}{16\pi^2}
\left(-\frac{N_c}{12\pi^2F}\right)
\left(\frac{8\pi^2F^2}{N_c M_{V_1}^2}\right)\\
&\times\bigg\{
\frac{ M_{V_1}^2}{2m_e^2}\big[B_0(m_e^2,m_e^2, M_{V_1}^2)-B_0(0, M_{V_1}^2, M_{V_1}^2)-1\big]\\
&-\bigg(\frac{ M_{V_1}^2}{M_\pi^2}-1\bigg)\big[B_0(m_e^2,m_e^2, M_{V_1}^2)-B_0(0,m_e^2,m_e^2)-2\big]\\
&+C^\prime_0(m_e^2,m_e^2,M_\pi^2,0,m_e^2, M_{V_1}^2)\\
&-C^\prime_0(m_e^2,m_e^2,M_\pi^2, M_{V_1}^2,m_e^2, M_{V_1}^2)
\bigg\}\,,
\end{split}
\end{equation}
\begin{align*}
&P_{\pi^0\to e^+e^-}^\text{THS}
=-\frac{e^4m_e}{16\pi^2}
\left(-\frac{N_c}{12\pi^2F}\right)\\
&\times\Bigg\{
-\frac{(2\pi F)^2}{N_c}\bigg\{\frac{1}{m_e^2}+\frac{24}{M_\pi^2}
+\bigg[\frac{ M_{V_1}^2}{( M_{V_2}^2- M_{V_1}^2)}\\
&\times\bigg(\frac1{m_e^2}\big[B_0(m_e^2,m_e^2, M_{V_1}^2)-B_0(0, M_{V_1}^2, M_{V_1}^2)\big]\\
&+\frac2{M_\pi^2}\big[7\,B_0(m_e^2,m_e^2, M_{V_2}^2)-B_0(m_e^2,m_e^2, M_{V_1}^2)\\
&-6\,B_0(0,m_e^2,m_e^2)\big]\bigg)\bigg]+\bigg[ M_{V_1}\leftrightarrow M_{V_2}\bigg]\\
&-\frac{2}{ M_{V_2}^2- M_{V_1}^2}\bigg(1+\frac{2\kappa M_{V_1}^4 M_{V_2}^4}{(4\pi F)^6M_\pi^2}\bigg)\\
&\times\big[B_0(m_e^2,m_e^2, M_{V_2}^2)-B_0(m_e^2,m_e^2, M_{V_1}^2)\big]\bigg\}\\
&+C^\prime_0(m_e^2,m_e^2,M_\pi^2,0,m_e^2,0)\\
&-\bigg\{\frac{ M_{V_2}^2}{ M_{V_2}^2- M_{V_1}^2}\bigg(2-\frac{3}{N_c}\frac{(4\pi F)^2}{ M_{V_2}^2}\bigg)\\
&\times C^\prime_0(m_e^2,m_e^2,M_\pi^2,0,m_e^2, M_{V_1}^2)-\frac{ M_{V_2}^4}{2( M_{V_2}^2- M_{V_1}^2)^2}\\
&\times\bigg[2+\frac{\kappa M_{V_1}^4}{N_c(4\pi F)^4}-\frac{(4\pi F)^2}{N_c M_{V_2}^2}\bigg(6+\frac{ M_{V_1}^2}{ M_{V_2}^2}\bigg)\bigg]\\
&\times C^\prime_0(m_e^2,m_e^2,M_\pi^2, M_{V_1}^2,m_e^2, M_{V_1}^2)\bigg\}-\bigg\{ M_{V_1}\leftrightarrow M_{V_2}\bigg\}\\
&-\frac{ M_{V_1}^2 M_{V_2}^2}{( M_{V_2}^2- M_{V_1}^2)^2}\\
&\times\bigg[2+\frac{\kappa M_{V_1}^2 M_{V_2}^2}{N_c(4\pi F)^4}-\frac{14}{N_c}(2\pi F)^2\bigg(\frac1{ M_{V_1}^2}+\frac1{ M_{V_2}^2}\bigg)\bigg]\\
&\times C^\prime_0(m_e^2,m_e^2,M_\pi^2, M_{V_1}^2,m_e^2, M_{V_2}^2)
\Bigg\}\,.
\tag{\stepcounter{equation}\theequation}
\end{align*}


\providecommand{\href}[2]{#2}\begingroup\raggedright\endgroup


\end{document}